\documentclass[aps,prl,article,twocolumn,superscriptaddress]{revtex4-1}

\usepackage{graphicx}
\usepackage{natbib,aas_macros}
\usepackage{hyperref}
\usepackage{color}
\hypersetup{colorlinks=true,linkcolor=blue,filecolor=blue,urlcolor=blue,citecolor=blue}

\begin{document}

\title{High-energy neutrinos and gamma rays from winds and tori in active galactic nuclei}

\author{Susumu Inoue}
\email[]{sinoue@chiba-u.jp}
\affiliation{International Center for Hadron Astrophysics, Chiba University, Chiba, Japan}
\affiliation{Faculty of Education, Bunkyo University, Koshigaya, Japan}
\affiliation{Astrophysical Big Bang Laboratory and Interdisciplinary Theoretical and Mathematical Sciences Program (iTHEMS), RIKEN, Wako, Japan}

\author{Matteo Cerruti}
\email[]{cerruti@apc.in2p3.fr}
\affiliation{Universit\'e Paris Cit\'e, CNRS, Astroparticule et Cosmologie, F-75013 Paris, France}
\affiliation{Universitat de Barcelona, ICCUB, IEEC-UB, E-08028 Barcelona, Spain}

\author{Kohta Murase}
\email[]{murase@psu.edu}
\affiliation{Department of Physics; Department of Astronomy \& Astrophysics, Center for Multimessenger Astrophysics of Institute for Gravitation and the Cosmos, Penn State University, University Park, USA}
\affiliation{School of Natural Sciences, Institute for Advanced Study, Princeton, New Jersey 08540, USA}
\affiliation{Yukawa Institute of Theoretical Physics, Kyoto University, Kyoto, Japan}

\author{Ruo-Yu Liu}
\email[]{ryliu@nju.edu.cn}
\affiliation{Nanjing University, Nanjing, China}

\date{\today}

\begin{abstract}
Powerful winds with wide opening angles, likely driven by accretion disks around black holes (BHs), are observed in the majority of active galactic nuclei (AGN) and can play a crucial role in AGN and galaxy evolution.
If protons are accelerated in the wind near the BH via diffusive shock acceleration, $pp$ and $p\gamma$ processes
generate neutrinos as well as pair cascade emission from the gamma-ray to radio bands.
The TeV neutrinos detected by IceCube from the obscured Seyfert galaxy NGC 1068
may arise from collisionless shocks in a failed, line-driven wind that is physically well motivated.
Although the cascade emission is $\gamma\gamma$-attenuated above a few MeV,
it can still contribute significantly to the sub-GeV gamma rays and the sub-millimeter emission observed from NGC 1068.
At higher energies, gamma rays can occur via $pp$ processes from a shock where an outgoing wind impacts the obscuring torus, along with some observable GHz-band emission.
Tests and implications of this model are discussed.
Neutrinos and gamma rays may offer unique probes of AGN wind launching sites,
particularly for objects obscured in other forms of radiation.
\end{abstract}

\maketitle


Active galactic nuclei (AGN) are likely powered mainly by accretion disks around supermassive black holes (BHs) \cite{Krolik99}.
Less than 10\% of all AGN in the present Universe are of the radio-loud class that produce 
powerful, collimated jets of plasma with ultra-relativistic outflow velocities \cite{Blandford19}.
The majority of AGN are instead classified as radio-quiet and do not possess prominent jets. 
Nonetheless, there is widespread evidence that most AGN can eject winds of thermal plasma
with wide opening angles ($2 \theta_w \gtrsim 60-100$ deg)
and a range of outflow velocities ($v_w \sim$ few $100 {\rm \, km \, s^{-1}}$ - $0.4 c$),
observable as blue-shifted atomic absorption features in the ultraviolet (UV) to X-ray bands \cite{Veilleux05, King15, Laha21}.
The fastest winds in X-rays are known as ultrafast outflows (UFOs) and
seen in $\gtrsim$40\% of all nearby AGN, of both radio-loud and radio-quiet types \cite{Tombesi13, Tombesi14}.
Inferred to occur on sub-pc scales,
their kinetic power can reach a substantial fraction 
of the bolometric luminosity $L_{\rm bol}$ \cite{Gofford15, Nardini18}.
In relatively nearby AGN, winds can also be discerned in the kinematics of their narrow emission line regions (NLR) on sub-kpc scales \cite{Veilleux05}
\footnote{See Supplemental Material at [URL] for more details on AGN winds, formulation, analytic estimates, results and discussion}.

Such AGN-driven winds may be launched from accretion disks by mechanisms involving radiative and/or magnetic processes \cite{Ohsuga14, Laha21}.
Winds are potentially ubiquitous in AGN with sufficiently high Eddington parameter $\lambda_{\rm Edd}$
(ratio of AGN bolometric luminosity $L_{\rm bol}$ to Eddington luminosity \cite{Frank02})
\footnote{
Objects that so far lack evidence of winds may simply have unfavorable viewing angles \cite{Nomura16, Giustini19} and/or have not been observed when the variable absorption features are visible \cite{Igo20}.}.
AGN winds may play crucial roles in the collimation of jets in radio-loud AGN \cite{Blandford19}, as well as in the evolution of supermassive BHs and their host galaxies through their feedback effects onto their environment \cite{Fabian12, Harrison18, Note1}.

The kinetic energy of AGN winds may be partly dissipated and channelled into high-energy electrons and protons via mechanisms such as diffusive shock acceleration (DSA) \cite{Blandford87, Bell13}.
This can induce non-thermal emission, e.g. from external shocks
where the wind interacts with the host galaxy gas \cite{Faucher12, Wang15, Lamastra16, Liu18}.
Despite some tentative evidence \cite{Zakamska14, Panessa19}, such emission is yet to be clearly discerned.

Also likely generic to all relatively luminous AGN is a geometrically thick torus of dusty, clumpy gas surrounding the nucleus on pc scales \cite{Netzer15, RamosAlmeida17}.
Depending on its inclination relative to the observer, such tori can substantially absorb the optical to X-ray emission from the accretion disk, resulting in the known differences between type-1 (unobscured) and type-2 (obscured) AGN.
The absorbed energy is reprocessed into the observed infrared (IR) emission
\footnote{The torus may naturally form within the inflow of gas toward the BH due to radiation pressure and other effects \cite{Netzer15, Wada16}.}.

NGC 1068, an archetypal type-2 Seyfert galaxy at distance $D \sim 10-16 {\rm \, Mpc}$ \cite{BlandHawthorn97},
is a known source of GeV gamma rays \cite{FermiLAT12} as well as TeV neutrinos \cite{IceCube20, IceCube22}.
Although UV-X-ray signatures of winds on sub-pc scales are unobservable due to high obscuration by its torus \cite{Bauer15, Marinucci16, Zaino20}, its NLR on larger scales exhibit an outflow with $v_{w, {\rm NLR}} \lesssim 2000 {\rm \, km \, s^{-1}}$
and $L_{w, {\rm NLR}} \lesssim 10^{43} {\rm \, erg \, s^{-1}}$, likely driven by the accretion disk \cite{Cecil90, MuellerSanchez11, GarciaBurillo14, GarciaBurillo19, Revalski21}.
The nature of the gamma rays detected by Fermi-LAT at energies $E_\gamma \sim 0.1-20 {\rm \, GeV}$ \cite{FermiLAT12} is unclear,
exceeding the inferred level associated with star formation in the host galaxy
(i.e. $pp$ $\pi^0$-decay gamma rays from interaction of cosmic rays from supernovae and interstellar gas) \cite{YoastHull14, Eichmann16}
\footnote{See however \cite{Ajello23}.}.

Neutrino observations by IceCube \cite{IceCube22} reveal that the most significant position in the northern hemisphere in a full-sky scan is coincident with that of NGC 1068.
Independently, a 4.2$\sigma$ excess over background expectations is found at its position in a source catalog search.
The spectrum is quite soft, with muon neutrino flux best fit as $f_{\nu_\mu} \propto \varepsilon_\nu^{-3.2}$
at energies $\varepsilon_\nu \sim 1.5-15 {\rm \, TeV}$,
and inferred luminosity $\varepsilon_\nu dL_{\nu_\mu}/d\varepsilon_\nu \sim 3 \times 10^{42} {\rm \, erg \, s^{-1}}$ in this range.
Meanwhile, upper limits for gamma rays above 0.2 TeV
\cite{MAGIC19}
rule out models in which TeV gamma rays and neutrinos escape the source with similar flux
\footnote{These include models involving kpc-scale wind external shocks \cite{Lamastra16}}.
Some recent proposals invoke proton acceleration and neutrino production in hot coronal regions near the BH where X-rays are emitted via thermal Comptonization, either accretion disk coronae \cite{Murase20, Kheirandish21} or accretion shocks \cite{YInoue19, YInoue20},
so that accompanying gamma rays would be significantly absorbed via $\gamma\gamma$ interactions with AGN photons
\cite{Murase22, Note1}.

Here we propose an alternative picture where protons are accelerated in the inner regions of the wind relatively near the BH in NGC 1068, which has various advantages over the coronal region models \cite{Note1}.
DSA, a well established mechanism for particle acceleration, is assumed.
This region may be identified with a ``failed'' wind that is plausibly expected in radiative, line-driven wind models for the conditions corresponding to NGC 1068 \cite{Giustini19}.
Neutrinos are mainly generated via $pp$ and $p\gamma$ interactions in this region,
while $\gamma\gamma$ interactions mediate the associated pair cascade emission, which we evaluate across the full EM spectrum.
For the GeV gamma rays, we invoke a separate region where the wind interacts with the torus, accelerates protons via DSA and induces $pp$ interactions with the torus gas.
This allows GeV photons to escape, while TeV photons are $\gamma\gamma$-absorbed by IR photons from the torus.
All relevant emission processes are modeled self-consistently with a detailed numerical code.
We use the notation $X_a=X/10^a$ for normalized variables.

{\it \bf Formulation}.
DSA at collisionless shock waves with sufficiently high Mach numbers can convey a sizable fraction of the energy of bulk plasma motion into that of non-thermal particles \cite{Blandford87, Bell13}.
In the inner regions of AGN winds near the BH, 
shocks may naturally form \cite{Sim10} in failed winds that are robustly expected in models of line-driven winds from the accretion disk
\cite{Murray95a, Proga00, Proga04, Nomura16, Nomura20}, 
particularly for the BH mass $M_{\rm BH}$ and $\lambda_{\rm Edd}$ inferred for NGC 1068 \cite{Giustini19, Note1}. 
Such flows are initially launched from the inner parts of the disk (typically at radii $R \lesssim 100 R_s$, where $R_s = 2 G M_{\rm BH}/c^2$ is the Schwarzschild radius),
but do not exceed the escape velocity $v_{\rm esc} = (2 G M_{\rm BH}/R)^{1/2}$ due to overionization \cite{Note1}
and eventually fall back,
thereby interacting with gas flowing out subsequently .
Henceforth we assume that protons are accelerated by DSA in the inner regions of the wind,
with the total proton power $L_p$ as a parameter.

\begin{figure}[htb!]
\includegraphics[width=1.1\linewidth]{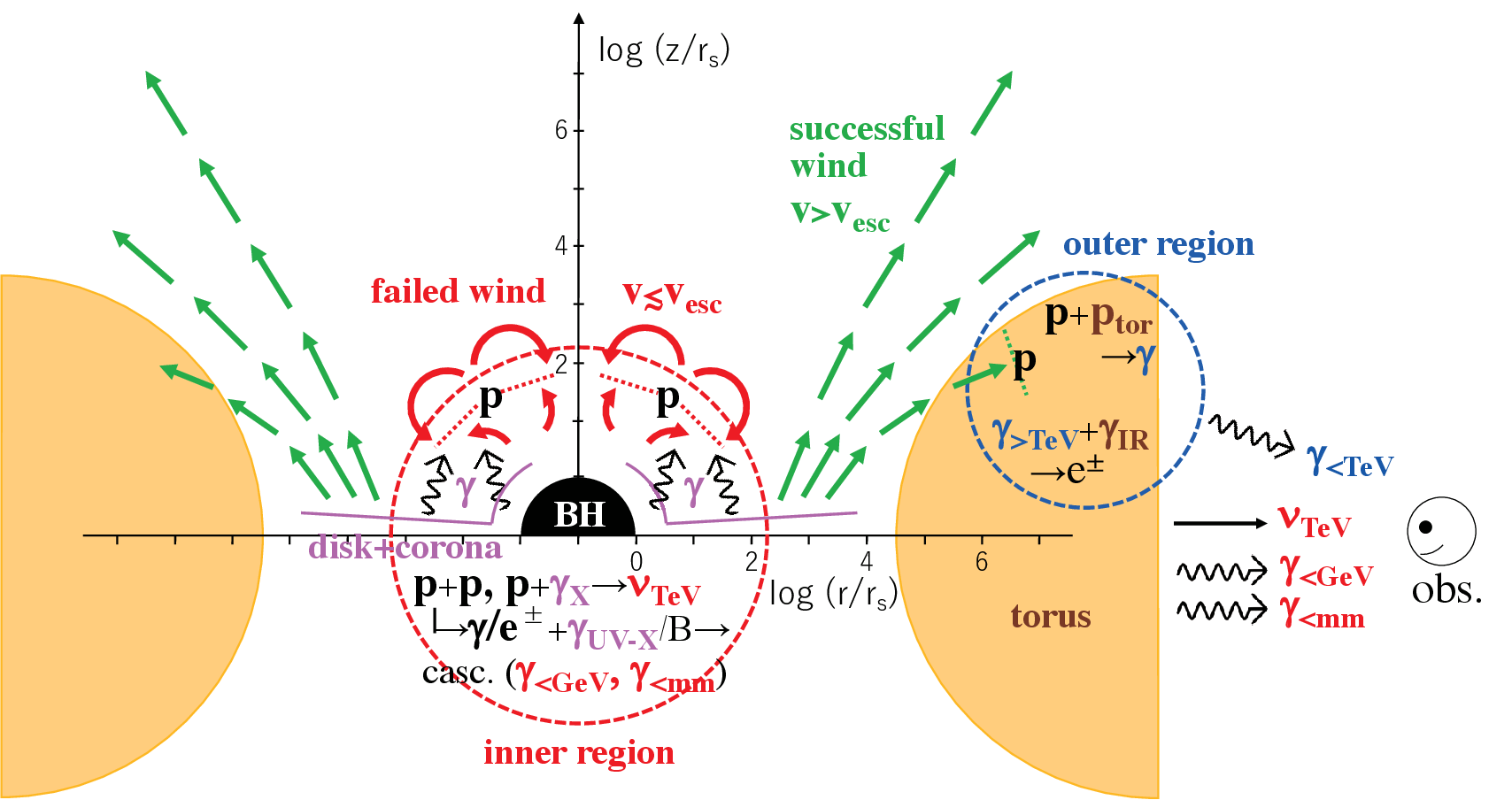}
\caption{
Schematic sketch of the model.
The accretion disk around the black hole (BH) drives an outflowing wind.
Inner region: winds from the inner disk dissipate their kinetic energy via shocks near the BH,
caused by failed line-driven winds that fall back.
Protons undergo diffusive shock acceleration (DSA) and $pp$ ($p\gamma$) interactions with gas (photons from the disk and corona), inducing neutrino and electromagnetic cascade emission, modulated by $\gamma\gamma$ interactions. 
Outer region: successful winds from the outer disk
propagate farther, partially impact the torus and trigger shocks.
Protons undergo DSA and $pp$ interactions with the torus gas, inducing gamma-ray emission, affected by $\gamma\gamma$ interactions with photons from the torus.
Indicated scales are only approximate.}
\label{fig:sketch}
\end{figure}

At the same time, a successful wind exceeding $v_{\rm esc}$ can be line-driven from the outer parts of the disk, mainly in the equatorial direction that is shielded from ionization \cite{Note1}.
This outer wind can propagate farther and impact the torus \cite{GarciaBurillo19, Bannikova22},
potentially inducing strong shocks and DSA of protons \cite{Mou21b}, for which we assume a total proton power $L_{p,o}$.
The model geometry is illustrated in Fig.~\ref{fig:sketch}.

Employing a numerical code that builds on previous work (LeHa-Paris) \cite{Cerruti15, Cerruti21}, we model the multi-messenger (MM) emission induced by a population of high-energy protons interacting with magnetic fields, gas and/or radiation \cite{Note1}.
For either the inner region of the failed wind or the outer region of the wind-torus interaction,
the emission region is a uniform, stationary sphere of radius $R_x$ with gas density $n_x$ and tangled magnetic field of amplitude $B_x$, through which all charged particles are advected with bulk flow velocity $v_x$. 
The index $x$ is denoted $o$ for the outer region, while it is dropped when referring to the inner region.

For both regions, $pp$ interactions between accelerated protons and ambient gas are important
\footnote{We also approximately account for the contribution of nuclei heavier than helium \cite{Note1}.
For simplicity, the terms ``protons'' and ``$pp$'' are used even when including nuclei.}.
The inner region is also permeated by radiation from the AGN that is the dominant target for $p\gamma$ and $\gamma\gamma$ interactions as well as seed photons for inverse Compton (IC) processes
\footnote{Hadronic interactions between photons and nuclei heavier than helium are neglected, as the role of $p\gamma$ processes for MM emission turns out to be mostly sub-dominant compared to $pp$ processes in this work.}.

\begin{figure}[htb!]
\includegraphics[width=1.1\linewidth]{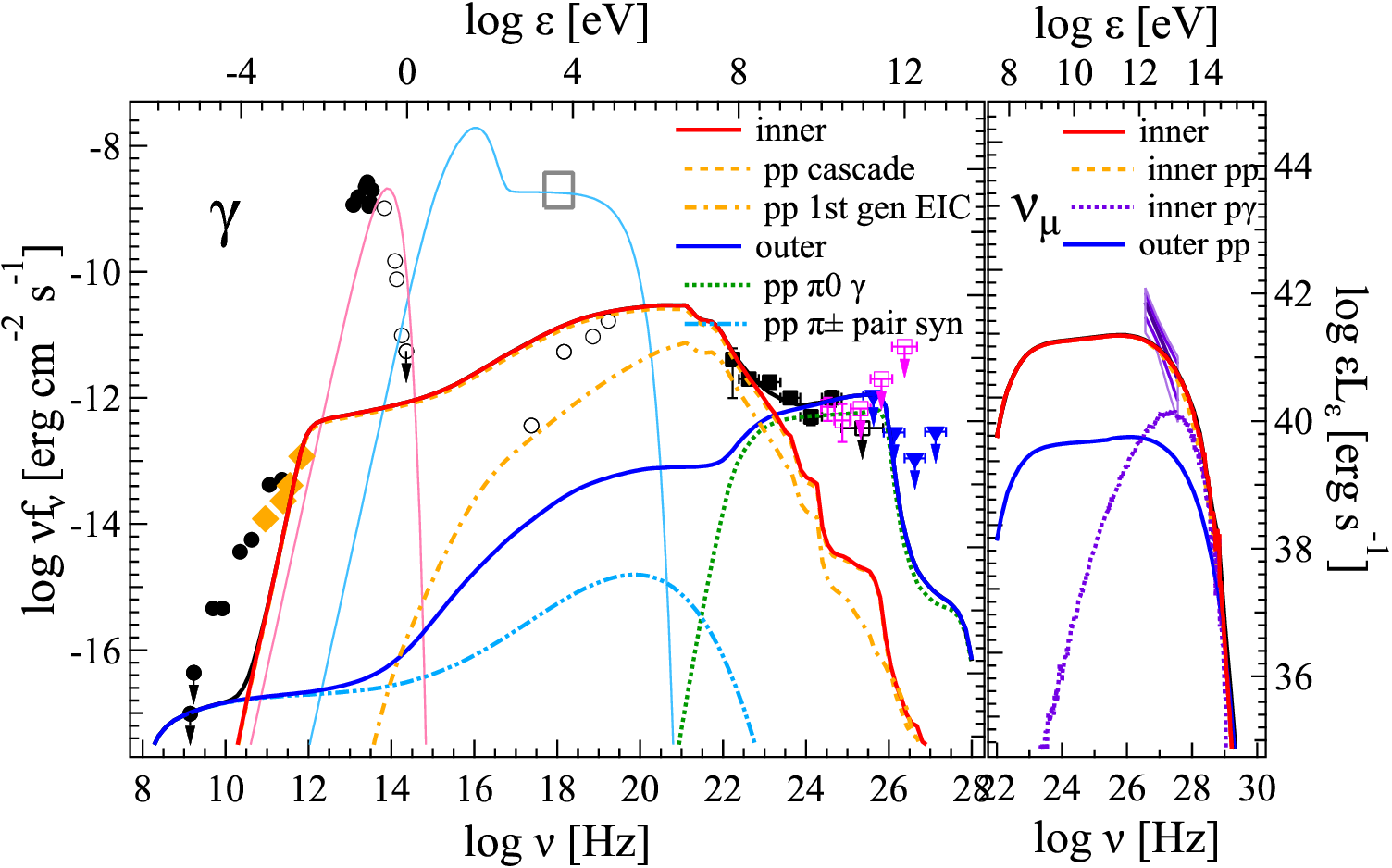}
\caption{
Model vs. observations of the multi-messenger spectrum of NGC 1068
for fiducial parameters.
Inner region: $L_p=7.1 \times 10^{43} {\rm erg \, s^{-1}}$, $R=30 R_s$ ($n=8.7 \times 10^{9} \ {\rm cm^{-3}}$), $\epsilon_B=0.5$ ($B=129 {\rm \, G}$), $\eta_g=3770$.
Outer region: $L_{p,o} = 1.3 \times 10^{42} {\rm \, erg \, s^{-1}}$, $R_o=0.1 {\rm \, pc}$, $n_o = 10^6 {\rm \, cm^{-3}}$, $B_o=15 {\rm \, mG}$.
Total emission from the inner (red solid), outer (blue solid), and both (black solid) regions shown.
Left: Electromagnetic spectrum.
Components dominating each band highlighted:
total $pp$ cascade (ochre dashed), 
external inverse Compton (EIC) from first-generation $pp$ cascade pairs (ochre dot-dashed), 
$pp$ $\pi^0$ decay (green dotted),
$pp$ $\pi^\pm$ decay pair synchrotron (cyan double-dot-dashed).
Assumed disk+corona (cyan thin) and torus (magenta thin) components overlaid.
Data plotted for radio to X-rays on sub-pc scales \cite{Prieto10} (black circles),
distingushing bands affected by obscuration (empty circles),
high resolution ALMA (ochre diamonds) \cite{Michiyama23}, 
Fermi-LAT \cite{FermiLAT17, FermiLAT20} (black and magenta squares) and MAGIC \cite{MAGIC19} (blue triangles).
Intrinsic X-ray flux (gray box) indicated \cite{Marinucci16}.
Right: Muon neutrino spectrum. Inner $pp$ (ochre dashed) and $p\gamma$ (purple dotted) highlighted.
Best fit line (thick), 1- (medium) and 2- (light) $\sigma$ error regions from IceCube denoted \cite{IceCube22}.
}
\label{fig:SED_fid}
\end{figure}

Adopting $D = 14 {\rm \, Mpc}$,
its spectrum is of a standard, geometrically thin accretion disk \cite{Shakura73, Frank02}
around a BH with $M_{\rm BH}=3 \times 10^7 M_\odot$ \cite{Greenhill96, Gallimore96_maser},
peaking in the optical-UV at $\varepsilon_{\rm disk} \simeq 32 {\rm \, eV}$
with total luminosity $L_{\rm disk} \simeq L_{\rm bol} =10^{45} {\rm \, erg \, s^{-1}}$ \cite{Woo02} (implying $\lambda_{\rm Edd} \simeq 0.27$),
plus an X-ray emitting corona with photon index $\Gamma_{\rm cor}=2$,
exponential cutoff energy $\varepsilon_{\rm cor} = 128 {\rm \, keV}$ \cite{Bauer15}
and 2-10 keV luminosity $L_{\rm cor, 2-10} = 7 \times 10^{43} {\rm \, erg \, s^{-1}}$ \cite{Marinucci16},
adopting parameters consistent with observations of NGC 1068 \cite{Note1} (Fig.~\ref{fig:SED_fid}).

For the inner region, we assume $v=v_{\rm esc}=c(R/R_s)^{-1/2}$,
as expected for failed winds.
In steady state, the bulk of the plasma should escape the region via advection
on a dynamical timescale $t_{\rm dyn}=R/v$,
either from the polar regions toward the BH,
or from the equatorial regions toward the accretion disk
\footnote{In addition, a minor fraction of thermal particles with velocities higher than the average value $v \sim v_{\rm esc}$
may become unbound and escape outwards.}.
As the average pre-shock velocity of particles is $v \sim v_{\rm esc}$,
the total pre-shock kinetic energy 
$E_k \sim (1/2) m_p n v_{\rm esc}^2 (4 \pi/3) R^3 = (2 \pi m_p / 3) n R^3 v_{\rm esc}^2$.
If a fraction $\epsilon_p$ of this is channelled into accelerated protons,
$\epsilon_p E_k \sim L_p t_{\rm dyn}$,
so that
$n = 3 L_p t_{\rm dyn} / 2 \pi m_p \epsilon_p R^3 v_{\rm esc}^2 = 3 L_p / 2 \pi m_p \epsilon_p R^2 v_{\rm esc}^3$,
thus relating $n$ with $R$, $L_p$ and $\epsilon_p$.

The outer region is the part of the torus that interacts strongly with the outer wind,
mostly where the latter grazes the torus funnel, rather than near the disk where gas must be inflowing \cite{GarciaBurillo19} (Fig.~\ref{fig:sketch}).
Besides gas as $pp$ targets,
it is immersed in thermal radiation that serves mainly as $\gamma\gamma$ targets,
emitted from the inner torus of radius $R_{\rm tor}=0.1 {\rm \, pc} \gtrsim R_o$
and temperature $T_{\rm tor} = 1000 {\rm \, K}$, as constrained by near IR observations of NGC 1068
\cite{GarciaBurillo19, GamezRosas22, Note1}.
We also set $v_o = 5000 {\rm \, km \, s^{-1}}$ for physical consistency with an outer, successful wind
in scenarios of line-driven winds \cite{Giustini19}.

For each region, protons are injected with total power $L_{p,x}$ and a power-law distribution in energy,
$dN_p/dE_p \propto E_p^{-2}$,
from $E_{p,\min}=m_p c^2$ up to an exponential cutoff characterized by $E_{p,\max,x}$
\footnote{For simplicity, injection of primary electrons is not considered in this work.}.
Subsequent $pp$ and/or $p\gamma$ (photopion and photopair or Bethe-Heitler; BeH) interactions lead to production of secondary hadrons, leptons and photons, of which charged pions and muons decay into neutrinos.
Photons with sufficient energy trigger pair cascades via $\gamma\gamma$ interactions.
The charged particles generate photons by synchrotron and IC processes.
The steady state distribution of all particles and the resulting MM emission are obtained self-consistently
by solving the coupled kinetic equations that account for their radiative losses and advective escape \cite{Note1}
\footnote{The treatment of energy losses on the proton distribution is not fully numerical and only approximate \cite{Note1}.}.
As $v_x$ is at most mildly relativistic,
Doppler effects are weak, and the emission is quasi-isotropic 
\footnote{This refers to the emission before obscuration by the torus.}.

The value of $E_{p,\max,x}$ is set where
DSA is limited by the available time or radiative losses,
$t_{{\rm acc},x}(E_p)=\min [t_{{\rm dyn},x}, t_{{\rm rad},x}(E_p)]$,
where $t_{{\rm acc},x}(E_p)=(10/3) (c \eta_{g,x} E_p /  e B_x v_{r,x}^2)$ is the DSA timescale
and $\eta_{g,x} \gtrsim 1$ parameterizes the strength of magnetic turbulence
\footnote{More precisely, what enters in the expression for $t_{{\rm acc}}$ should be the shock velocity $v_s$ that can differ from the flow velocity $v$ by a factor of order unity, but we make the simplification $v_s \sim v$}, 
$t_{{\rm dyn},x}=R_x/v_{r,x}$ 
and $t_{{\rm rad},x}(E_p)$ is the radiative loss timescale,
primarily $pp$, photopion and BeH losses for the inner region,
$t_{\rm rad}=(t_{pp}^{-1}+t_{p\gamma\pi}^{-1}+t_{{\rm BeH}}^{-1})^{-1}$,
and $pp$ loss for the outer region,
$t_{{\rm rad},o}=t_{pp}$.

We posit $\epsilon_p = 0.2$, a reasonable value for DSA.
For $B$, $B^2/8\pi=\epsilon_B (1/2) m_p n v^2$ 
with $\epsilon_B \lesssim 0.5$ delineates a plausible range.
The main parameters are
$L_p$, $R$, $\epsilon_B$, $\eta_g$ for the inner region 
and $L_{p,o}$, $R_o$, $n_o$, $B_o$ for the outer region
\footnote{The parameters $v_o$ and $\eta_{g,o}$ are not critical for the results; see below}.
These are adjusted to best describe the MM data, 
also considering physical plausibility.

{\it \bf Analytic estimates}.
First we conduct analytic estimates 
that justify our choice of parameters
and are broadly consistent with our numerical results (Fig.~\ref{fig:timescale}; \ref{fig:SED_fid}, right panel). 
For either $pp$ or $p\gamma$ processes,
protons with $E_p$ induce neutrinos with $\varepsilon_\nu \sim 0.05 E_p$,
implying $E_p \sim 40 {\rm \, TeV} (\varepsilon_\nu/ 2 {\rm \, TeV})$.
For the inner region,
$n = (3 L_p / 2 \pi m_p c^3 \epsilon_p R_s^2) \bar{R}^{-1/2}
\simeq 0.87 \times 10^{10} {\rm \, cm^{-3}} \epsilon_{p, -0.7}^{-1} L_{p,43.85} \bar{R}_{1.5}^{-1/2}$
where $\bar{R}=R/R_s$, $R_s = 0.89 \times 10^{13} {\rm \, cm}$, $\bar{R}_{1.5} = \bar{R}/30$,
$\epsilon_{p, -0.7} = \epsilon_p/0.2$ and $L_{p,43.85} = L_p / 7.1 \times 10^{43} {\rm \, erg \, s^{-1}}$,
hence the $pp$ loss timescale
$t_{pp} 
\sim 1.7 \times 10^5 {\rm \, s} \ \epsilon_{p, -0.7} L_{p,43.85}^{-1} \bar{R}_{1.5}^{1/2}$ at $E_p \gg 1 {\rm \, GeV}$
\cite{Aharonian04}
\footnote{Although the numerical value here is for $E_p = 40 {\rm \, TeV}$,
the $pp$ cross section increases only logarithmically with $E_p$
when sufficiently above the threshold at $E_p \sim 1 {\rm \, GeV}$ \cite{Aharonian04}.}.

For our assumed disk+corona radiation field,
the total $p\gamma$ loss timescale is roughly
$t_{\rm p\gamma} = (t_{p\gamma\pi}^{-1} + t_{\rm BeH}^{-1})^{-1}
\sim 1.2 \times 10^6 {\rm \, s} \, \bar{R}_{1.5}^2 (E_p / 40 {\rm \, TeV})^{-1}$ (Fig. \ref{fig:timescale}) \cite{Note1}.
With $v=v_{\rm esc} \simeq 5.5 \times 10^5 {\rm \, km \, s^{-1}} \, \bar{R}_{1.5}^{-1/2}$
and $B=(6 \epsilon_B L_p /\epsilon_p r^2 v)^{1/2}
\simeq 129 {\rm \, G} \, \epsilon_{B, -0.3}^{1/2} \epsilon_{p, -0.7}^{-1/2} L_{p, 43.85}^{1/2} \bar{R}_{1.5}^{-3/2}$
where $\epsilon_{B, -0.3} = \epsilon_B/0.5$,
$t_{\rm acc} \simeq
3.4 {\rm \, s} \, \eta_g \, \epsilon_{B, -0.3}^{-1/2} \epsilon_{p, -0.7}^{1/2} L_{p, 43.85}^{-1/2} \bar{R}_{1.5}^{5/2} (E_p / 40 {\rm \, TeV})$.
If $\eta_g$ is $\sim 1-40$ as inferred for DSA in supernova remnants \cite{Bell13},
$E_{p,\max}
\sim 24 {\rm \, PeV}\, \eta_g^{-1/2} \epsilon_{B, -0.3}^{1/4} \epsilon_{p, -0.7}^{-1/4} L_{p, 43.85}^{1/4} \bar{R}_{1.5}^{-1/4}$,
limited by $t_{\rm p\gamma}$ (Fig. \ref{fig:timescale}). 
In this case, $p\gamma$ neutrinos would dominate with a hard spectrum peaking in power at a few PeV,
unlike what is acually seen in NGC 1068. 

The observed soft spectrum at $\varepsilon_\nu \sim 2 - 20 {\rm \, TeV}$ favors a much lower $E_{p,\max}$,
which can be realized if $\eta_g \gg 1$, as inferred for DSA in the jets of blazar AGN \cite{Inoue96}. 
As long as $\eta_g
\gtrsim 560 \, \epsilon_{B, -0.3}^{1/2} \epsilon_{p, -0.7}^{-1/2} L_{p, 43.85}^{1/2} \bar{R}_{1.5}^{-3/2}$,
 $t_{\rm dyn} 
\simeq 4.9 \times 10^4 {\rm \, s} \, \bar{R}_{1.5}^{3/2}$ limits $E_{p,\max}$
so that the maximum neutrino energy
$\varepsilon_{\nu,\max} = 0.05 E_{p,\max}
\sim 7.5 {\rm \, TeV}\, (\eta_g/3770)^{-1} \epsilon_{B, -0.3}^{1/2} \epsilon_{p, -0.7}^{-1/2} L_{p, 43.85}^{1/2} \bar{R}_{1.5}^{-1}$.
As $t_{pp} \lesssim  t_{p\gamma}$ for $E_p \lesssim
290 {\rm \, TeV} \epsilon_{p, -0.7}^{-1} L_{p, 43.85} \bar{R}_{1.5}^{3/2}$,
$pp$ neutrinos dominate over $p\gamma$.

Considering mixing among the three neutrino flavors during propagation,
the $pp$ neutrino luminosity per flavor per log $\varepsilon_\nu$ is 
${\cal L}_{\nu_\mu} (\varepsilon_\nu)
\approx (1/6) X_{\rm nuc} f_{pp, {\rm net}}(E_p=20 \varepsilon_\nu) {\cal L}_p (E_p=20 \varepsilon_\nu)$,
where ${\cal L}_p (E_p)$
is the injected proton power per log $E_p$,
$f_{pp, {\rm net}} = t_{pp}^{-1}( t_{pp}^{-1}+t_{p\gamma}^{-1}+t_{\rm dyn}^{-1})^{-1}$
is the net $pp$ efficiency \cite{Murase16}, 
and $X_{\rm nuc} \simeq 2.0$ is a factor accounting for the effect of nuclei \cite{Note1}.
When $E_{p,\max}$ is low enough that $t_{pp} \ll t_{p\gamma}$,
the weak $E_p$-dependence of $t_{pp}$ implies
${\cal L}_{\nu_\mu} (\varepsilon_\nu)
\propto \varepsilon_\nu^0 \exp(-\varepsilon_\nu/\varepsilon_{\nu,\max})$
($f_{\nu_\mu} \propto \varepsilon_\nu^{-2} \exp(-\varepsilon_\nu/\varepsilon_{\nu,\max})$),
roughly similar to the spectrum of injected protons.
A suitable $\varepsilon_{\nu,\max}$ can provide a reasonable account of the observed soft neutrino spectrum.

If $t_{\rm dyn} \lesssim t_{pp}$,
$f_{pp, {\rm net}} \sim t_{\rm dyn}/t_{pp}
\simeq 0.3 \, \epsilon_{p, -0.7}^{-1} L_{p, 43.85} \bar{R}_{1.5}$,
whereas if $t_{\rm dyn} \gtrsim t_{pp}$, $f_{pp, {\rm net}} \simeq 1$.
Approximating ${\cal L}_p \sim L_{p} \ln (E_{p,\max}/E_{p,\min})$,
to reproduce ${\cal L}_{\nu_\mu}(\varepsilon_\nu = 2 {\rm \, TeV}) \sim 5 \times 10^{41} {\rm \, erg \, s^{-1}}$ as observed,
$L_p \sim 7.1 \times 10^{43} {\rm \, erg \, s^{-1}} \bar{R}_{1.5}^{-1/2}$
\footnote{Note that from our assumption relating $n$ with the pre-shock kinetic energy,
$f_{pp,{\rm net}}$ itself depends on $L_p$.},
which is somewhat optimistic but within a plausible range. 
As $f_{pp, {\rm net}} \propto R$ when $f_{pp, {\rm net}} <1$, 
smaller $R$ would demand still higher $L_p$.
On the other hand, larger $R$ can make
the Thomson optical depth of the inner region
$\tau_e = \sigma_T \mu_e n R
\simeq 1.8 \, \epsilon_{p, -0.7}^{-1} L_{p,43.85} \bar{R}_{1.5}^{1/2}$
too high to be consistent with a line-driven wind
\footnote{Moreover, $\tau_e  \gg 1$ could make the shocks radiation-mediated and suppress DSA,
as well as prevent penetration of low-energy photons for effective $\gamma\gamma$ absorption \cite{Note1}}.
These arguments constrain $R$, $L_p$
\footnote{These constraints on $R$ and $L_p$ are also consistent with the shocks being collisionless and conducive to DSA \cite{Note1}.}
and the combination $\eta_g \epsilon_B^{-1/2}$.
Comparing the numerically calculated EM cascade with radio observations further constrains $\epsilon_B$ \cite{Note1}.

The optical depth for $\gamma\gamma$ interactions with corona photons 
$\tau_{\gamma\gamma, {\rm cor}}(\varepsilon)
\simeq 140 \, (\varepsilon/{\rm 1 \, GeV}) \, \bar{R}_{1.5}^{-1}$
\cite{Note1},
so gamma rays co-produced with neutrinos are attenuated above several MeV, similar to the coronal region models.

For the outer torus region,
$E_{p,\max,o} \sim 460 {\rm \, TeV} \eta_{g,o}^{-1} (B_o/ 1 {\rm \, mG})(R_o/0.1 {\rm \, pc})$,
limited by $t_{{\rm dyn}, o} = R_o/v_o \simeq 6.2 \times 10^8 {\rm \, s} (R_o/0.1 {\rm \, pc})$.
The $pp$ gamma-ray luminosity per log $\varepsilon$ is 
${\cal L}_{\gamma} (\varepsilon)
\approx (1/3) X_{\rm nuc} f_{pp, {\rm net}, o} {\cal L}_{p, o} (E_p=10 \varepsilon)$ \cite{Note1}
where the pp loss timescale
$t_{pp, o}
\simeq 1.6 \times 10^9 {\rm \, s} (n_o/10^6 {\rm \, cm^{-3}})^{-1}$.
Reproducing ${\cal L}_{\gamma}(\varepsilon = 1 {\rm \, GeV}) \sim 3 \times 10^{40} {\rm \, erg \, s^{-1}}$ as observed \cite{FermiLAT12}
is feasible with e.g.
$R_o = 0.1 {\rm \, pc}$, $n_o = 10^6 {\rm \, cm^{-3}}$ and
$L_{p, o} \sim 3 f_{pp, {\rm net}, o}^{-1} {\cal L}_{\gamma} \ln(10^6) \sim 2.5 \times 10^{42} (f_{pp, {\rm net}, o}/0.5)^{-1}  {\rm \, erg \, s^{-1}}$.
An ambient blackbody radiation field with $T_{\rm tor}=1000 {\rm \, K}$ and $R_{\rm tor} =0.1 {\rm \, pc}$
implies $\tau_{\gamma\gamma, {\rm tor}} \gtrsim 1$ for $\varepsilon \gtrsim 0.2 {\rm \, TeV}$ \cite{Note1},
consistent with TeV observations \cite{MAGIC19}
\footnote{As gamma rays corresponding to $E_{p,\max,o}$ are unobservable,
$v_o$ and $\eta_{g,o}$ are not crucial.
However, $B_o \lesssim 15$ mG is required for consistency with radio observations.}.

{\it \bf Numerical results}.
Numerical calculations generally confirm our analytic estimates for the neutrinos, and also allow detailed studies of the broadband EM emission caused by complex hadronic cascade processes. 
Guided by the estimates above,
we fiducially adopt for the inner region
$L_p = 7.1 \times 10^{43} {\rm \, erg \, s^{-1}}$, 
$R=30 R_s \simeq 2.7 \times 10^{14} {\rm \, cm}$,
$\epsilon_B = 0.5$ ($B = 129 {\rm \, G}$) 
and $\eta_g = 3770$. 
These parameters imply $E_{p,\max} \simeq 150 {\rm \, TeV}$ and $n=8.7 \times 10^{9} \ {\rm cm^{-3}}$. 
Other values of $R$, $\epsilon_B$ and $\eta_g$ are also studied
\footnote{See Fig.\ref{fig:SED_etaBcomp} and Fig.\ref{fig:SED_Rcomp} for results with non-fiducial values of $R$, $\epsilon_B$ and $\eta_g$ \cite{Note1}}.
For the outer region,
we choose $R_o = 0.1 {\rm \, pc}$ and $n_o = 10^6 {\rm \, cm^{-3}}$,
and adjust $B_o$ and $L_{p,o}$ to be consistent with the EM data.

Fig.\ref{fig:SED_fid} presents the fiducial numerical results compared with the available MM data for NGC 1068.
Neutrinos from the inner region are predominantly due to $pp$ but also contain a non-negligible $p\gamma$ component,
with spectral cutoffs at $\varepsilon_{\nu, \max} \sim$ 7.5 TeV, generally consistent with the current IceCube data.
Also present is a sub-dominant contribution of $pp$ neutrinos from the outer region.

EM emission from the inner region is dominated by the $pp$ cascade 
\footnote{See Fig.\ref{fig:SED_inner} for a detailed description of different emission components for the inner region \cite{Note1}.}.
Despite considerable $\gamma\gamma$ attenuation above several MeV as expected, 
it is luminous enough to contribute substantially to the observed sub-GeV emission, 
mostly via IC upscattering of AGN photons by first-generation $pp$ cascade pairs 
\footnote{This was also seen but not clearly emphasized for the coronal region models \cite{Murase20, YInoue20}.
See also Ref. \cite{Ajello23, Murase24}.}.
At higher energies, $pp$ gamma rays from the outer region take over, where $L_{p,o} = 1.3 \times 10^{42} {\rm \, erg \, s^{-1}}$
\footnote{See Fig.\ref{fig:SED_outer} for a detailed description of different emission components for the outer region \cite{Note1}.}.
Above $\sim$0.1 TeV, the $pp$ gamma rays are severely $\gamma\gamma$-attenuated
by the torus IR radiation, in agreement with the current upper limits.

The cascade emission from the inner region extends down to the radio-far IR bands,
and can contribute significantly to the sub-mm emission detected by ALMA on sub-pc scales \cite{YInoue20, Michiyama23}
as long as $\epsilon_B \gtrsim 0.1$ \cite{Note1}.
Also observationally relevant may be GHz-band synchrotron emission from the outer region
by secondary pairs from $pp$-induced $\pi^{\pm}$ decay
\footnote{The general origin of radio emission in radio-quiet AGN is under debate,
one possibility being non-thermal emission from AGN winds \cite{Panessa19}.
Here we suggest a specific emission site on sub-pc scales in the context of winds, 
although it would only explain a part of the radio spectrum of NGC 1068.}.
For consistency with the current upper limit at a few GHz, we choose $B_o = 15 {\rm \, mG}$, 
within the range inferred from independent polarization measurements for the inner torus of NGC 1068 \cite{LopezRodriguez15}.
This implies $E_{p, \max, o} \simeq 6 \, \eta_{g, o}^{-1} {\rm \, PeV}$, set by $t_{{\rm acc}, o}=t_{\rm dyn, o}$.

{\it \bf Caveats.}
The principal assumption here for the inner region is that
near the BH at $R \sim 30 R_s$,
protons are accelerated on a timescale
$t_{\rm acc} 
\sim \eta_g c E_p /e B v_{\rm esc}^2$ 
with total power $L_p \lesssim 0.1 L_{\rm bol}$
and undergo $pp$ interactions with ambient protons on a timescale $t_{\rm dyn} \sim R/v_{\rm esc}$. 
Plausibly accounting for the neutrinos observed from NGC 1068 
entails $\eta_g \sim$ a few 1000. 
We identify this with regions of failed winds
that are robustly expected in line-driven models of AGN winds \cite{Note1}.

Although shocks in failed, line-driven winds are seen in numerical simulations \cite{Sim10},
they are yet to be analyzed in detail.
Failed winds have been proposed to be the origin of some other known features of AGN,
such as dense X-ray obscurers near the nucleus (possibly present in NGC 1068 \cite{Zaino20}),
part of the broad line region, and the soft X-ray excess
\cite{Giustini19, Giustini21, Note1}.
Further studies, both theoretical and observational, are warranted
to clarify the existence and properties of failed winds in AGN, and how they compare with the current model.

For the outer region, future work should account for the clumpy structure of the torus, 
synchrotron and IC emission from primary electrons, as well as an underlying starburst contribution \cite{Note1}.

{\it \bf Tests and implications.}
Similar to the coronal region models \cite{Murase20, Kheirandish21, YInoue19, YInoue20},
the prominent MeV-GeV cascade emission (Fig.~\ref{fig:SED_fid}) may be decipherable with current GeV and/or future MeV-GeV instruments \cite{Engel22} for NGC 1068 and other nearby AGN.
As the sub-GeV emission arises from $R \sim 30 R_s$, 
variability is expected on timescales down to hours,
albeit with limited amplitude due to weak Doppler effects.
MM variability correlations between neutrinos, sub-GeV and/or optical emission on longer timescales caused by variations in the mass accretion rate ${\dot M}$ provide a potential test \cite{Note1}.
The $\gamma\gamma$ origin of the TeV gamma-ray break can be tested with CTA \cite{CTA19}.

The significant sub-mm cascade emission is a unique feature 
that not only constrains $\epsilon_B$
but also allows distinction with other models.
Compared to $B \sim 100 {\rm \, G}$ here,
magnetic fields 
in the disk corona \cite{Murase20, Kheirandish21} and accretion shock \cite{YInoue19, YInoue20} models
are $\sim 1000 {\rm \, G}$ and $\sim 10 {\rm \, G}$, respectively.
In the former, the cascade may not reach the sub-mm band 
due to stronger synchrotron self-absorption (SSA),
while in the latter, it may be much less luminous and possibly dominated by synchrotron from primary electrons.
Due to the SSA break at $\sim$THz,
the spectrum of the cascade alone may be flatter than observed below a few 100 GHz (Fig.~\ref{fig:SED_fid})
\footnote{Nevertheless, accounting for the majority of the observed sub-mm emission with the cascade may also be possible
by considering a suitable radial distribution of physical properties, beyond our one zone model \cite{Note1}}.
Additional emission from the dusty torus, small-scale jet, etc. could be important in this band \cite{Michiyama23}.
Variability and MM correlations similar to the sub-GeV emission, as well as unresolved morphology in future VLBI imaging by ngVLA \cite{Reid18},
should help distentangle the cascade from other components \cite{Note1}.

Although detailed discussions are beyond the current scope, we may speculate on expectations of this model for AGN other than NGC 1068.
In coronal region models, the neutrino luminosity would correlate with the X-ray luminosity, and the fact that NGC 1068 is the brightest AGN for IceCube is attributed to its intrinsic X-ray brightness \cite{Murase20,YInoue20}, combined with its favorable declination for the detector \cite{Kheirandish21}.
In the current wind model, an additional factor may be high $\lambda_{\rm Edd}$,
required for high wind power \cite{Giustini19, Nomura20} (also valid for the Circinus galaxy \cite{Note1}.)
The extent of the region of failed winds may depend systematically on $\lambda_{\rm Edd}$, $M_{\rm BH}$ and the UV-X-ray spectra of AGN \cite{Giustini19}.
This implies important differences from other models for the neutrino and cascade EM emission of different AGN,
as well as the diffuse neutrino background from all AGN, to be explored in future work.

Most intriguingly, neutrinos and gamma rays may be unique probes of the inner regions of AGN where winds are launched from the accretion disk and interact with their immediate environment, especially in obscured objects.
Future high-energy MM observations may provide important new insight into the physics of AGN winds,
which are widely believed to play a critical role in the evolution of supermassive BHs and galaxies.

\acknowledgments
We thank Mariko Nomura, Ken Ohsuga, Keiichi Wada, Yoshiyuki Inoue, Shigeo Kimura and Aya Ishihara for valuable discussions.
S.I. is supported by KAKENHI Nos. 17K05460 and 20H01950. M.C. received financial support through the Postdoctoral Junior Leader Fellowship Programme from la Caixa Banking Foundation, grant No. LCF/BQ/LI18/11630012. 
K.M. is supported by NSF Grant Nos. AST-1908689, AST-2108466 and AST-2108467, and KAKENHI Nos. 20H01901 and 20H05852. R.-Y.L. is supported by NSFC Grant No. U2031105.
\bibliographystyle{apsrev4-2}
\bibliography{Awind-gamnu}

\clearpage

\section{Supplemental Material}

\section{Overview of AGN winds}

There is widespread evidence that the majority of AGN can commonly eject winds of thermal plasma with wide opening angles and a range of outflow velocities, observable as blue-shifted atomic absorption features in the ultraviolet (UV) to X-ray bands \cite{Crenshaw03, Veilleux05, King15, Laha21}.
Those with velocities from a few 100 to a few 1000 ${\rm km \, s^{-1}}$ are known as warm absorbers (WAs) in X-rays and narrow absorption line outflows in the UV.
Those with still higher velocities, up to mildly relativistic values of $\sim 0.4 c$, are called ultrafast outflows (UFOs) in X-rays and broad absorption line (BAL) outflows in the UV.
BAL outflows are observed in $\sim$20\% of quasars, primarily of the radio-quiet type \cite{Bruni14}.
UFOs are seen in $\sim$40\% of all nearby AGN, of both radio-loud and radio-quiet types \cite{Tombesi13, Tombesi14}.
WAs are detected in $\sim$65\% of nearby AGN, albeit being relatively rarer in radio-loud objects.
The opening angles of such winds are estimated to be $2 \theta_w \gtrsim$60-100 deg \cite{Veilleux05, Nardini15},
much wider than jets.

AGN winds may be launched from accretion disks by various mechanisms involving thermal, radiative and/or magnetic processes \cite{Ohsuga14, Giustini19, Kazanas19}.
UFOs and WAs may correspond to winds ejected from the disk at different ranges of radii \cite{Tombesi13, Kazanas19}.

AGN winds may be crucial for the collimation of relativistic jets in radio-loud AGN \cite{Tombesi12_jet, Blandford19}.
They may also be the primary conduits through which supermassive BHs exert feedback onto their host galaxies,
resulting in the known BH-galaxy scaling relations,
and possibly the quenching of star formation in massive galaxies
\cite{Fabian12, King15, Harrison18}.
Massive outflows of atomic and molecular material from AGN observed on kpc scales may be a manifestation of such effects \cite{Veilleux20}.

\section{Failed line-driven winds}

Although the actual mechanism for the formation of accretion-disk driven winds in AGN is not yet established,
one of the leading candidates is radiative, line-driven winds
\cite{Arav94, Murray95a, Proga00, Proga04, Risaliti10, Nomura16, Nomura17, Nomura20, Mizumoto21}.
Analogous to mechanisms discussed for winds from massive stars \cite{Castor75},
such winds utilize the enhanced pressure on gas by radiation at the frequencies of atomic line transitions, and are effective for high UV luminosities of the accretion disk.
However, simultaneously high X-ray luminosities of the disk corona can fully ionize the wind gas (``overionization'') and suppress momentum transfer in the inner parts of the disk, leading to outflows that are initially launched but fail to reach the escape velocity $v_{\rm esc}$ and eventually fall back toward the disk.
Such failed winds may be the origin \cite{Giustini21} of the dense X-ray obscurers seen in some type-1 AGN \cite{Kaastra14}
(possibly also in NGC 1068 \cite{Zaino20}), a part of the broad line region (BLR) on sub-pc scales \cite{Proga04}, or the soft X-ray excess common to many type-1 AGN \cite{Schurch06}.
Successful winds exceeding $v_{\rm esc}$ can still be driven from the outer parts of the disk that is shielded from ionizing X-rays, mainly in the equatorial direction.
The relative extent of the inner failed wind and the outer successful wind can vary with the UV to X-ray luminosity and spectrum,
which in turn depend on the BH mass $M_{\rm BH}$ and the Eddington parameter $\lambda_{\rm Edd}$.
For $M_{\rm BH}$ and $\lambda_{\rm Edd}$ inferred for NGC 1068, inner regions of failed winds are quite plausible \cite{Giustini19}.

Shocks can occur within the inner, failed wind region due to interactions among gas falling back and flowing out,
as indicated by some numerical simulations \cite{Sim10}.
At the same time,
the successful part of the wind can propagate farther and impact the torus  \cite{GarciaBurillo19, Bannikova22}, potentially inducing strong shocks in an outer region \cite{Mou21b}.

We note that the spatial scale $R$ of the failed wind region is not well constrained quantitatively, from either theory or observations.
Although a range of $R \lesssim 30-100 R_s$ has been invoked for numerical simulations
\cite{Proga00, Proga04, Nomura16, Nomura17, Nomura20},
explicit calculations of line-driven winds are currently limited to $> 30 R_s$ due to computational costs,
even with simplified treatments for the radiative transfer.
It also depends on whether the standard, thin accretion disk extends down to $\sim 3 R_s$
or is physically truncated at a larger radius, which is presently unclear for AGN.
Better constraints await further theoretical and observational developments.

\section{Formulation details}

The basic formulation for the model is overviewed in the main text.
The numerical treatment of photohadronic neutrino and cascade emission follows the LeHa-Paris code
as detailed in Refs. \cite{Cerruti15} and \cite{Cerruti21} and is not repeated here.
Here we provide some additional information not described elsewhere.

For $p\gamma$ and $\gamma\gamma$ interactions,
we always consider situations where external photons originating outside the emission region are dominant.
The calculations take into account
both synchrotron and inverse Compton (IC) radiation for the cascade emission, 
as well as Klein-Nishina loss effects according to Ref. \cite{Moderski05} for the equilibrium pair distribution.
For simplicity, the external radiation field is assumed to be uniform and isotropic throughout the region,
although more realistic calculations should account for its non-uniformity and anisotropy.

For $pp$ interactions,
the treatment of secondary photon, lepton, and neutrino production follows Ref. \cite{Kelner06}.
The contribution of nuclei heavier than helium for both accelerated hadrons and target gas
is approximately taken into account
by multiplying the production rate of all secondary particles in $pp$ interactions by the nuclear enhancement factor $X_{\rm nuc} = 2.0$ \cite{Mori09}.
This factor can be even higher if the abundance of heavy elements in the BH vicinity is super-solar,
as observationally inferred for some AGN \cite{Hamann99}.
The associated pair cascades are computed as for the $p\gamma$-induced cascades.

For the inner region, protons can suffer significant energy losses via $pp$ and $p\gamma$ interactions (Fig. \ref{fig:timescale}).
Instead of a fully numerical treatment,
the steady-state energy distribution of protons is calculated approximately as
\begin{eqnarray}
  \frac{dN_p(E_p)}{dE_p}
  &=& Q_p (E_p) \\ \nonumber
  &\times& \left( \frac{1}{\tau_{pp}(E_p)} + \frac{1}{\tau_{p\gamma\pi}(E_p)}  + \frac{1}{\tau_{BeH}(E_p)} + \frac{1}{\tau_{dyn}} \right)^{-1}
\label{eq:pdist}
\end{eqnarray}
where $Q_p(E_p) \propto E_p^{-2} \exp (-E_p/E_{p,\max})$ is the proton injection rate in units of cm$^{-3}$ s$^{-1}$.

Key assumptions for the inner region are
\begin{eqnarray}
  v = v_{\rm esc} = c \left(R \over R_s\right)^{-1/2} 
  \simeq 5.5 \times 10^5 {\rm \, km \, s^{-1}} \, \left(R \over 30 R_s\right)^{-1/2}
\label{eq:vesc}
\end{eqnarray}
and
\begin{eqnarray}
  n &=& {3 L_p t_{\rm dyn} \over 2 \pi m_p \epsilon_p R^3 v^2}
  = {3 L_p \over 2 \pi m_p \epsilon_p R^2 v^3} 
  \simeq 8.7 \times 10^{9} {\rm \, cm^{-3}} \\ \nonumber
  &\times& \left(\epsilon_p \over 0.2\right)^{-1} \left(L_p \over 7.1 \times 10^{43} {\rm \, erg \, s^{-1}}\right) \left(R \over 30 R_s\right)^{-1/2}
\label{eq:n_in}
\end{eqnarray}
that comes from the relation between the pre-shock kinetic energy and the energy in nonthermal protons.
The Thomson optical depth 
\begin{eqnarray}
  \tau_e &=& \sigma_T \mu_e n R \\ \nonumber
  &\simeq& 1.8 \left(\epsilon_p \over 0.2\right)^{-1} \left(L_p \over 7.1 \times 10^{43} {\rm \, erg \, s^{-1}}\right) \left(R \over 30 R_s\right)^{1/2} ,
\label{eq:tau_e}
\end{eqnarray}
where $\mu_e \sim 1.2$ is the electron mean molecular weight for fully ionized gas with solar abundance.
This should not be too high in order to
1) be consistent with line-driven winds,
2) avoid the shocks becoming radiation-mediated and suppressing DSA, which requires $\tau_e<c/v$ \cite{Levinson20}, and
3) allow permeation of low-energy photons for sufficient $\gamma\gamma$ absorption.
Low to moderate values of $\tau_e \sim 0.1-1$ are also implied for the coronal region models \cite{Murase20, YInoue20}.

The relaxation timescale via Coulomb collisions for thermal protons in the inner region is
\begin{eqnarray}
  t_{C,pp}
  &=& {4 \pi^{1/2} \over c \sigma_T} \left(m_p \over m_e\right)^2 {1 \over n \ln \Lambda} \left(k_B T_p \over m_p c^2\right)^{3/2} \\ \nonumber
  &=& {8 \pi^{3/2} m_p c^2 \over 3 \sigma_T} \left(m_p \over m_e\right)^2 \left(3 \mu \over 16\right)^{3/2} {1 \over \ln \Lambda} 
  {\epsilon_p R^2 v_{\rm esc}^6 \over L_p} \\ \nonumber
  &\simeq& 1.7 \times 10^6 {\rm \ s}
  \left(\epsilon_p \over 0.2\right) \left(L_p \over 7.1 \times 10^{44} {\rm \ erg \ s^{-1}} \right)^{-1} \left(R \over 30 R_s\right)^{-1} ,
\label{eq:tCpp_in}
\end{eqnarray}
where $\ln \Lambda \simeq 20$ is the Coulomb logarithm \cite{Murase20},
and the proton temperature is assumed to be $T_p = 3 \mu m_p v^2 / 16 k_B$ as appropriate for strong shocks,
with mean molecular weight $\mu \simeq 0.62$.
For the parameters considered here,
$t_{C,pp} \gg t_{\rm dyn} = R/v_{\rm esc} \simeq 4.9 \times 10^4 {\rm \, s} \, (R/30 \, R_s)^{3/2}$
so that the plasma is collisionless and accommodates DSA.

We provide details on the assumed external radiation fields,
which are crucial for the cascade as targets for $\gamma\gamma$ interactions and as seed photons for inverse Compton processes
(even if the products of $p\gamma$ interactions are sub-dominant compared to $pp$).
For the accretion disk,
\begin{eqnarray}
  \varepsilon \left(dL(\varepsilon) \over d\varepsilon\right)_{\rm disk}
   = L_{\rm disk,0} \left(\varepsilon \over \varepsilon_{\rm disk}\right)^{4/3} \exp \left(-{\varepsilon \over \varepsilon_{\rm disk}}\right) ,
\end{eqnarray}
where $\varepsilon_{\rm disk} =  2.82 k_B T_{\max}$, $T_{\max} = 0.488 T_{\rm in}$ is the maximum disk temperature,
and $T_{\rm in} = (G M_{BH} \dot{M}/72 \pi \sigma_{\rm SB} R_S^3)^{1/4}$
is the temperature at the innermost disk radius $r_{\rm in} = 3 R_S$ \cite{Frank02}.
Normalization is given by $L_{\rm disk, tot} = \int d\varepsilon (dL(\varepsilon)/d\varepsilon) = L_{\rm bol, obs}$.
Numerically,
\begin{eqnarray}
   \varepsilon_{\rm disk}=31.51 {\rm \ eV} \left(L_{\rm disk} \over 10^{45} {\rm erg\ s^{-1}}\right)^{1/4} \left(M_{\rm BH} \over 3 \times 10^7 M_\odot \right)^{-1/2}
\end{eqnarray}
and $L_{\rm disk,0}=1.12 L_{\rm disk}$.
We adopt $L_{\rm bol, obs} = 10^{45} {\rm erg\ s^{-1}}$ \cite{Woo02}.
The corresponding photon density in the emission region is
\begin{eqnarray}
   n_{\rm disk}(\varepsilon)
   &=& \left(dL(\varepsilon) \over d\varepsilon\right)_{\rm disk} {1 \over 4\pi c R^2} \\ \nonumber
   &=& {L_{\rm disk,0} \over 4\pi c R^2 \varepsilon_{\rm disk}}
\left(\varepsilon \over \varepsilon_{\rm disk}\right)^{1/3} \exp \left(-{\varepsilon \over \varepsilon_{\rm disk}}\right) ,
\end{eqnarray}
where
\begin{eqnarray}
   {L_{\rm disk,0} \over 4\pi c R^2 \varepsilon_{\rm disk}}
   &\simeq& 2.8 \times 10^{15} {\rm cm^{-3}} \\ \nonumber
   &\times& \left(\varepsilon_{\rm disk} \over 31.51 {\rm eV}\right)^{-1} \left(R \over 10 R_s \right)^{-2} \left(L_{\rm disk} \over 10^{45} {\rm erg\ s^{-1}}\right) .
\end{eqnarray}

If there is an outer radius $r_{\rm out}$ at which the disk is truncated,
the spectrum should transition from $\propto \varepsilon^{1/3}$ to $\propto \varepsilon^2$ below $\varepsilon = \varepsilon_{\rm out} = 2.82 k_B T_{\rm out}$,
where $T_{\rm out} = T_{\rm in} (r_{\rm out}/r_{\rm in})^{-3/4}$.
In the current model, this may effectively occur for $r \gtrsim R$, where the dilution with distance decreases their relative contribution to the emission region.
Numerically,
\begin{eqnarray}
   \varepsilon_{\rm out}=4.7 {\rm \ eV}
   \left(L_{\rm disk} \over 10^{45} {\rm erg\ s^{-1}}\right)^{1/4} \left(M_{\rm BH} \over 3 \times 10^7 M_\odot \right)^{-1/2} \\ \nonumber
 \times \left(R/r_{\rm in} \over 100/3 \right)^{-3/4} .
\end{eqnarray}

For the corona,
\begin{eqnarray}
   \varepsilon \left(dL(\varepsilon) \over d\varepsilon\right)_{\rm cor} = L_{\rm cor,0} \left(\varepsilon \over \varepsilon_0\right)^{2-\Gamma_{\rm cor}} \exp \left(-{\varepsilon \over \varepsilon_{\rm cor}}\right) ,
\end{eqnarray}
where $\varepsilon_0$ is a reference energy that can be chosen for convenience, e.g. $\varepsilon_0=2$ keV.
Normalization is given by the observed 2-10 keV luminosity $L_{\rm 2-10 keV, obs}$.
Since usually $\varepsilon_{\rm cor} \gg 10$ keV, 
for $\Gamma_{\rm cor}=2$,
$L_{\rm cor,0} = L_{\rm 2-10 keV, obs} / \ln (5)$,
whereas for $\Gamma_{\rm cor} \neq 2$,
$L_{\rm cor,0} = (2-\Gamma_{\rm cor}) L_{\rm 2-10 keV, obs} / (5^{2-\Gamma_{\rm cor}}-1)$.

Ref. \cite{Bauer15} give $\Gamma_{\rm cor} = 2.10^{+0.06}_{-0.07}$ and $\varepsilon_{\rm cor}=128^{+115}_{-44}$ keV.
Ref. \cite{Marinucci16} observed a putative unveiling event (temporary decrease of obscuring material along the line of sight), giving $L_{\rm 2-10 keV, obs} = 7^{+7}_{-3} \times 10^{43} {\rm erg\ s^{-1}}$.
We adopt $\Gamma_{\rm cor} = 2.0$, $\varepsilon_{\rm cor}=128$ keV (not directly relevant for our results as long as $\varepsilon_{\rm cor} \gg 10$ keV)
and $L_{\rm 2-10 keV, obs} = 7 \times 10^{43} {\rm erg\ s^{-1}}$,
which implies $L_{\rm cor,0} = 4.35 \times 10^{43} {\rm erg\ s^{-1}}$.
The corresponding photon density in the emission region is
\begin{eqnarray}
   n_{\rm cor}(\varepsilon)
   &=& \left(dL(\varepsilon) \over d\varepsilon\right)_{\rm cor} {1 \over 4\pi c R^2} \\ \nonumber
   &=& {L_{\rm cor,0} \over 4\pi c R^2 \varepsilon_0} \left(\varepsilon \over \varepsilon_0\right)^{1-\Gamma_{\rm cor}} \exp \left(-{\varepsilon \over \varepsilon_{\rm cor}}\right) ,
\end{eqnarray}
where
\begin{eqnarray}
   {L_{\rm cor,0} \over 4\pi c R^2 \varepsilon_0}
   &\simeq& 4.6 \times 10^{12} {\rm cm^{-3}} \\\nonumber
   &\times& \left(\varepsilon_0 \over 2 {\rm keV}\right)^{-1} \left(R \over 10 R_s \right)^{-2} \left(L_{\rm 2-10 keV, obs} \over 7 \times 10^{43} {\rm erg\ s^{-1}}\right) .
\end{eqnarray}

For the innermost region of the torus, a blackbody description is an adequate approximation for our purposes,
with fiducial radius $R_{\rm tor}=0.1$ pc and temperature $T_{\rm tor} =1000$ K \cite{GamezRosas22}.
The temperature alone gives the photon density,
\begin{eqnarray}
  n_{\rm tor}(\varepsilon) = {8\pi \over c^3 h^3} {\varepsilon^3 \over \exp(\varepsilon/k_B T_{\rm tor})-1}
\end{eqnarray}
As a numerical example, at $\varepsilon= 2.82 k_B T_{\rm tor} = 0.243 {\rm \ eV} (T_{\rm tor}/1000 K)$,
$n_{\rm tor}(\varepsilon=0.243 {\rm \ eV}) = 1.19 \times 10^{10} {\rm \ cm^{-3}}$.
The emission from both inner and outer regions are affected by $\gamma\gamma$ absorption with this radiation field.

Intergalactic gamma-ray attenuation due to $\gamma\gamma$ interactions with the extragalactic background light is included following Ref. \cite{Franceschini17},
although at $D=14$ Mpc, it is only a minor effect above a few tens of TeV.

\section{Additional results}

\begin{figure}[htb!]
\includegraphics[width=0.9\linewidth]{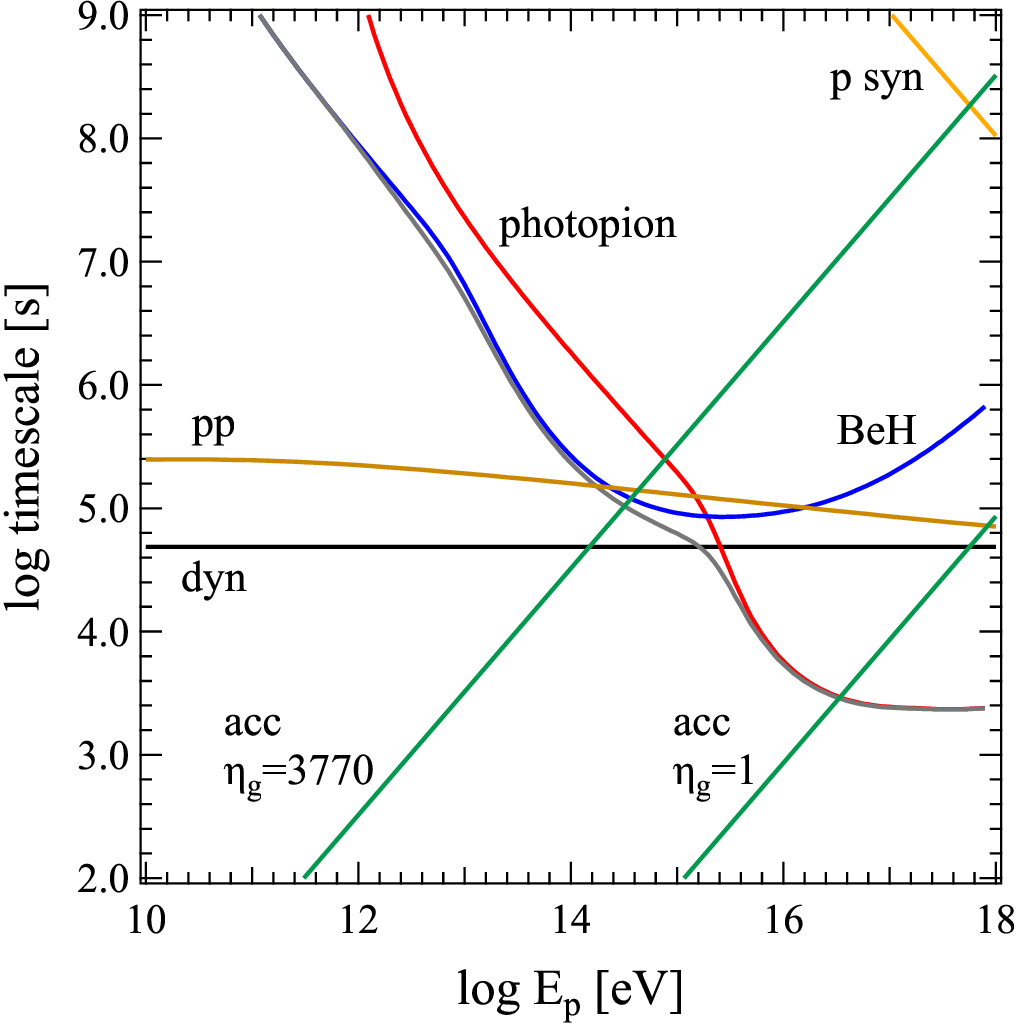}
\caption{
Comparison of timescales for the inner region in the fiducial case.
Dynamical time $t_{\rm dyn}$ (black), $pp$ loss time (brown), photopion loss time $t_{p\gamma\pi}$ (red), Bethe-Heitler loss time $t_{\rm BeH}$ (blue), total photohadronic loss time $t_{p\gamma}$ (gray),
proton synchrotron loss time $t_{\rm psyn}$ (ochre) \cite{Note1}
for $\epsilon_B=0.5$ ($B=129 {\rm \, G}$),
acceleration time $t_{\rm acc}$ (green) for \{$\epsilon_B=0.5$, $\eta_g=1$\} and \{$\epsilon_B=0.5$, $\eta_g=3770$\}.
}
\label{fig:timescale}
\end{figure}

Here we present more details on the fiducial results, as well as results for some non-fiducial range of parameters.
Fig.\ref{fig:timescale} compares the physically relevant timescales for the inner region in the fiducial case.
Comparison among $t_{\rm acc}$, $t_{\rm dyn}$ and $t_{\rm rad}$
determines $E_{p,\max}$, 
which would be $E_{p,\max} \sim 40 {\rm \, PeV}$ for $\eta_g = 1$, 
and $E_{p,\max} \sim 150 {\rm \, TeV}$ for $\eta_g = 3770$. 

\begin{figure}[htb!]
\includegraphics[width=1.0\linewidth]{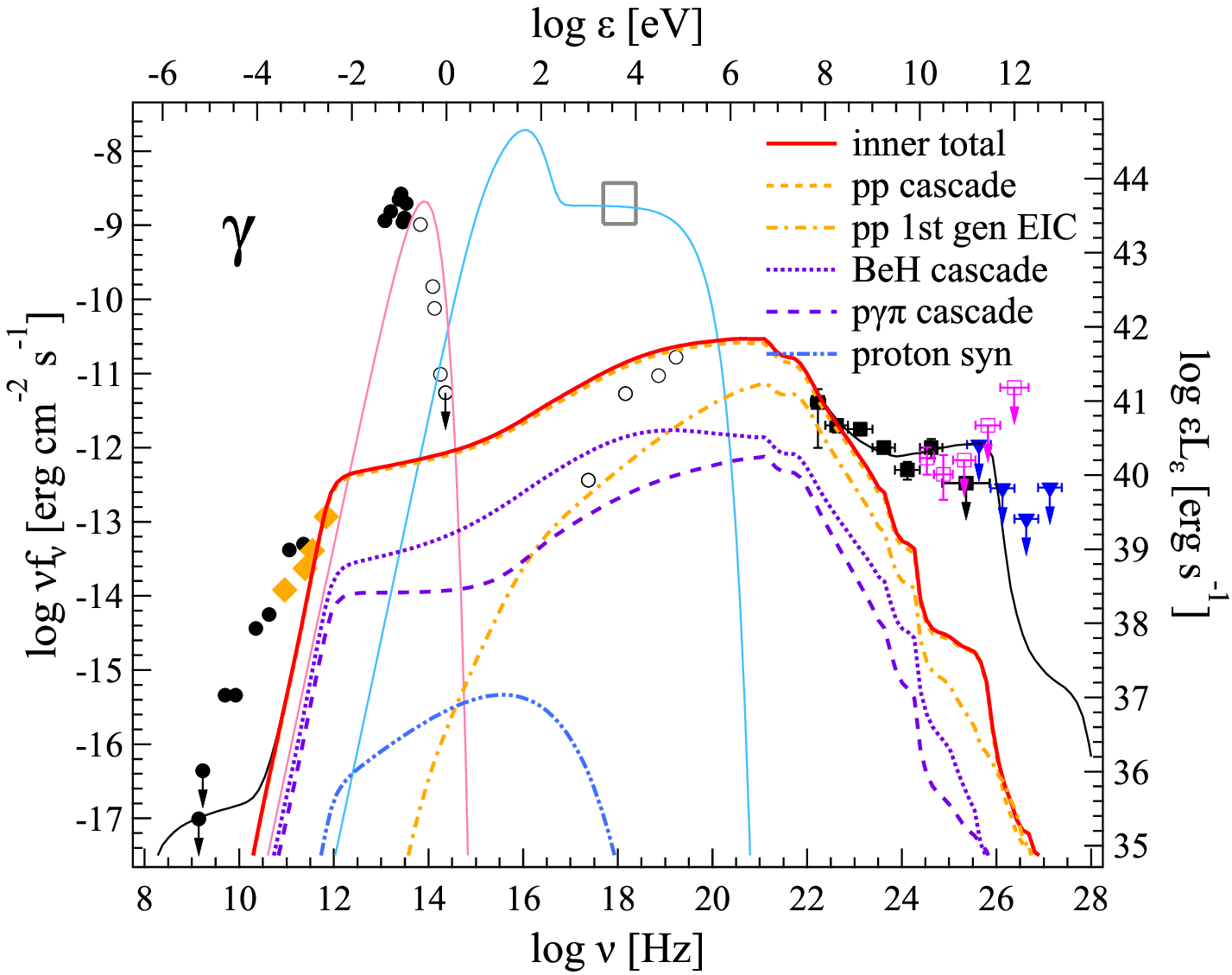}
\caption{
Model vs. observations of the electromagnetic spectrum of NGC 1068 for fiducial parameters,
clarifying the contribution of different emission components for the inner region, as indicated in the legend.
Total (red solid),
$pp$ cascade (ochre dashed), 
external inverse Compton (EIC) from first-generation $pp$ cascade pairs (ochre dot-dashed), 
$p\gamma$ Bethe-Heitler (BeH) cascade (purple short dashed),
photopion cascade (purple dashed),
proton synchrotron (blue double-dot-dashed).
Otherwise the same as the left panel of Fig.\ref{fig:SED_fid}.
}
\label{fig:SED_inner}
\end{figure}

\begin{figure}[htb!]
\includegraphics[width=1.0\linewidth]{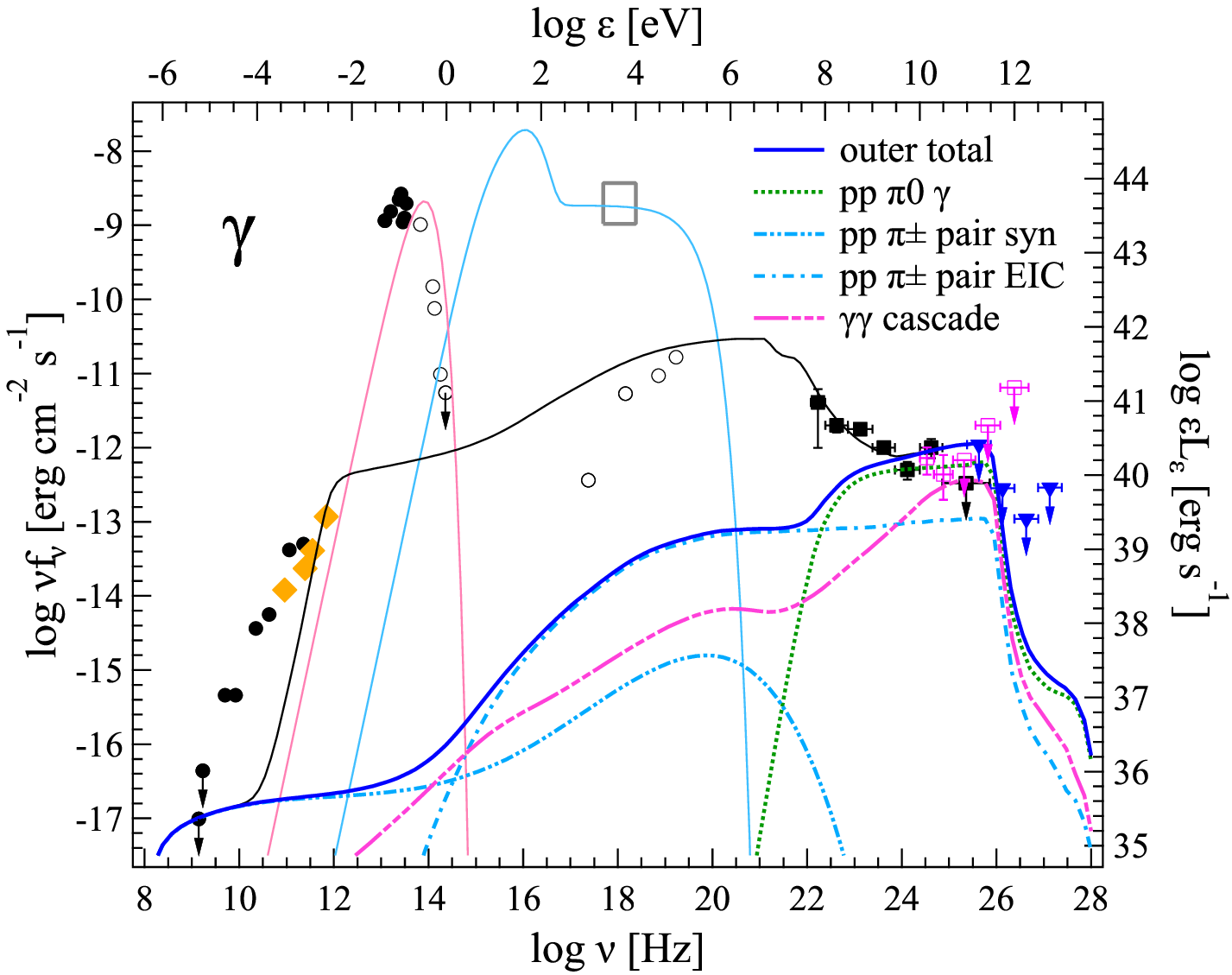}
\caption{
Model vs. observations of the electromagnetic spectrum of NGC 1068 for fiducial parameters,
clarifying the contribution of different emission components for the outer region, as indicated in the legend.
Total (blue solid),
$pp$ $\pi^0$ decay (green dotted),
$pp$ $\pi^\pm$ decay pair synchrotron (cyan double-dot-dashed),
$pp$ $\pi^\pm$ decay pair EIC (cyan dot-dashed),
$\gamma\gamma$ cascade (magenta triple-dot-dashed).
Otherwise the same as the left panel of Fig.\ref{fig:SED_fid}.
}
\label{fig:SED_outer}
\end{figure}

Figs.\ref{fig:SED_inner} and \ref{fig:SED_outer} clarify the separate contributions of different emission processes to the inner and outer regions, respectively, for the fiducial case.
For the inner region, $pp$ cascade dominates all bands, while Bethe-Heitler (BeH) and photopion cascade can be non-negligible in some bands. 
Proton synchrotron is subdominant.
For the outer region, each band is dominated by a different component.
Going from low to high energy:
$pp$ $\pi^\pm$ decay pair synchrotron, $\gamma\gamma$ cascade, $pp$ $\pi^\pm$ decay pair external inverse Compton (EIC), and $pp$ $\pi^0$ decay.

\begin{figure}[htb!]
\includegraphics[width=1.1\linewidth]{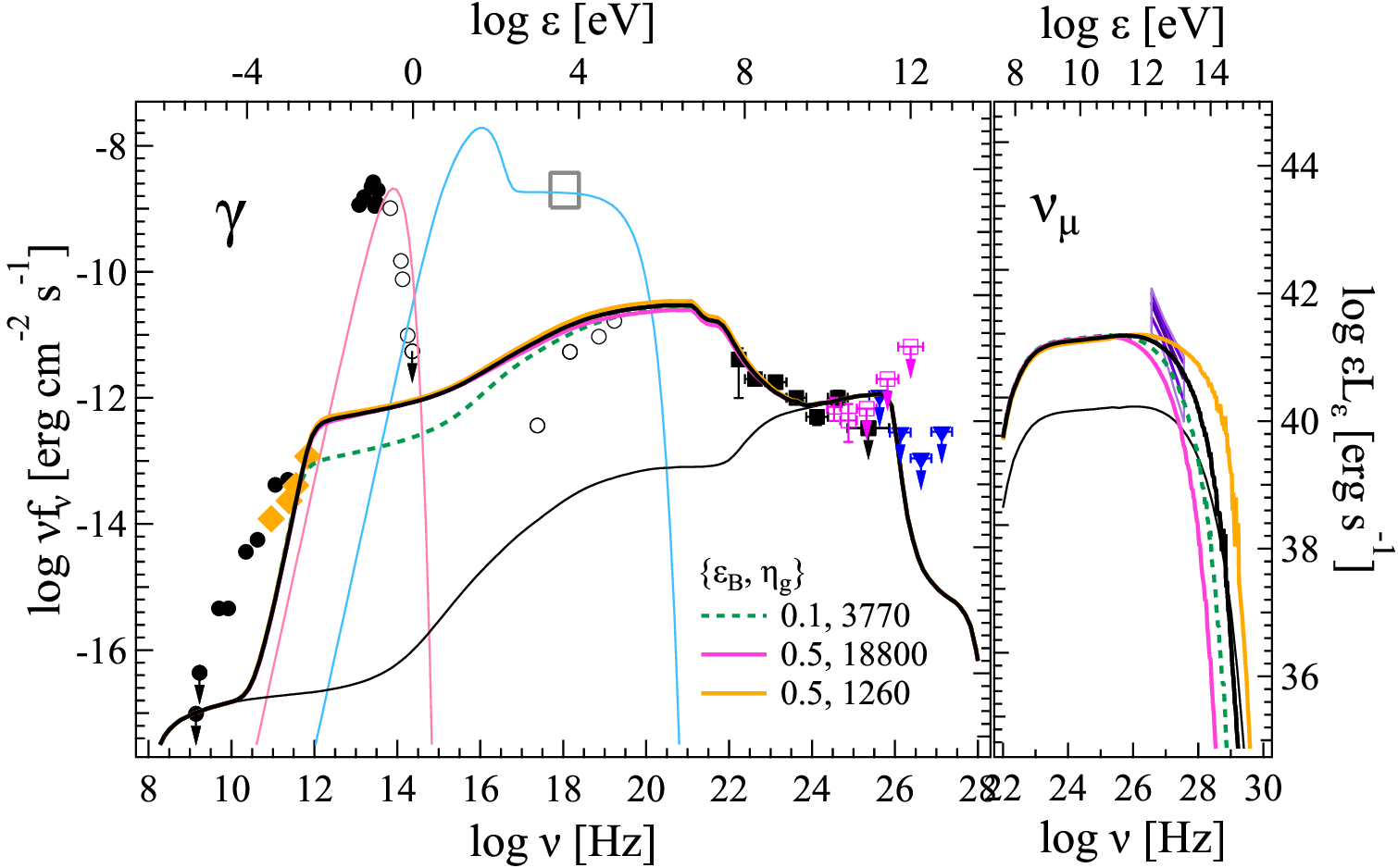}
\caption{
Model vs. observations of the multi-messenger spectrum of NGC 1068
for varying combinations of $\epsilon_B$ and $\eta_g$ for the inner region, as indicated in the legend.
Total emission from both regions shown for $\epsilon_B=0.5$ ($B=129 {\rm \, G}$) and $\eta_g=$1260, 3770 and 18800 (solid),
and $\epsilon_B=0.1$ ($B=57.8 {\rm \, G}$) and $\eta_g=$3770 (dashed).
Otherwise the same as Fig.\ref{fig:SED_fid}.
}
\label{fig:SED_etaBcomp}
\end{figure}

Fig.\ref{fig:SED_etaBcomp} shows cases with different combinations of $\epsilon_B$ and $\eta_g$ compared to the fiducial case.
To avoid the cascade emission exceeding the sub-millimeter emission detected by ALMA on scales $\lesssim$ 7 pc \cite{Michiyama23}, $\epsilon_B \lesssim 0.5$ is required (which is also in line with the physical requirement that the region should not be too magnetically-dominated to be congruent with a line-driven wind).
As long as $\epsilon_B \gtrsim 0.1$ ($B \gtrsim$ 60 G), the cascade can contribute significantly to the observed sub-mm emission, although the exact amount depends on the uncertain contributions from other emission components such as the dusty torus and the small-scale jet \cite{Michiyama23}.

The value of $\eta_g$ directly affects $E_{p,\max}$ and is mainly constrained by the higher energy end of the observed neutrino spectrum.
With other parameters being fiducial, a range of $\eta_g \sim$ 2000-10000 appears compatible with the current MM data.
With lower $\eta_g$ (higher $E_{p,\max}$), the contribution of $p\gamma$ neutrinos become more prominent (Fig. \ref{fig:timescale}).

\begin{figure}[htb!]
\includegraphics[width=1.1\linewidth]{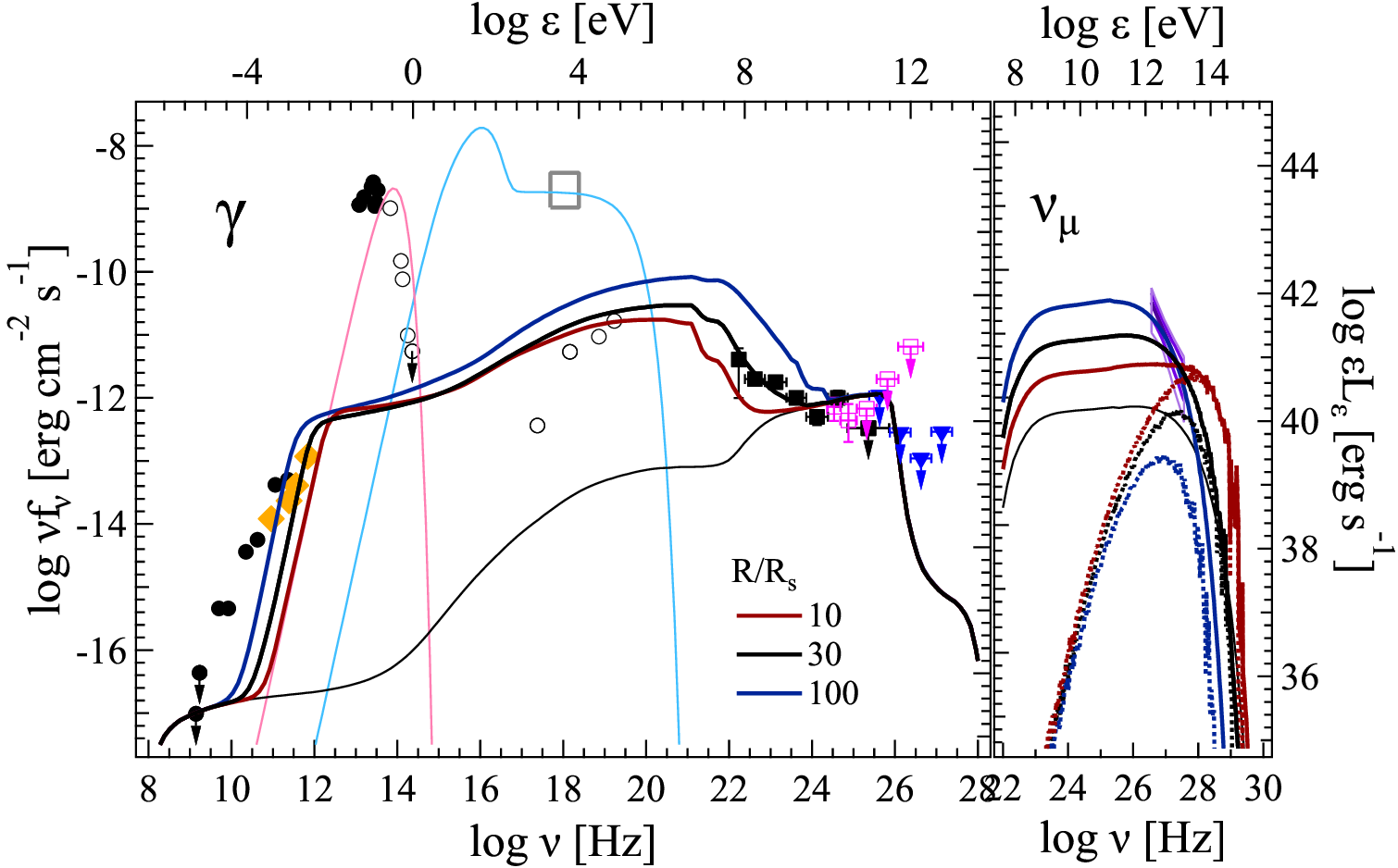}
\caption{
Model vs. observations of the multi-messenger spectrum of NGC 1068
when varying $R$ for the inner region.
Total emission from both regions shown for $R=10 R_s$ (dark red),
$R=30 R_s$ (fiducial, black)
and $R=100 R_s$ (dark blue),
along with total emission from the outer region (fiducial, thin black).
Contribution of p$\gamma$ neutrinos from the inner region are denoted separately (dashed).
Otherwise the same as Fig.\ref{fig:SED_fid}.
}
\label{fig:SED_Rcomp}
\end{figure}

Fig.\ref{fig:SED_Rcomp} compares the cases of $R/R_s=10$ and $R/R_s=100$ with the fiducial case.
Within this range of $R$,
the balance of $t_{\rm acc} \propto B^{-1} v^{-2} \propto R^{5/2}$
with $t_{\rm dyn} \propto R^{3/2}$
limit $E_{p,\max} \propto R^{-1}$ to $\sim$45 TeV for $R/R_s =10$ and $\sim$450 TeV for $R/R_s =100$.
(even though $t_{\rm rad} \propto R^2$ and $t_{pp} \propto R^{1/2}$
also become relatively important for $R/R_s =10$ and 100, respectively).
Whereas neutrino emission is completely dominated by $pp$ for $R/R_s=100$,
the relative contribution of $p\gamma$ becomes significant for $R/R_s=10$.
EM emission becomes more luminous with $R$ in bands affected by opacity,
for both $\gamma\gamma$ absorption at GeV and synchrotron self absorption at sub-mm.
Thus, although $R/R_s=100$ allows luminous neutrino emission,
it is disfavored due to overproduction of both GeV and sub-mm flux.
Conversely, $R/R_s=10$ has no such issues for the EM emission,
but the neutrino flux is underproduced.

We note that accounting for the broadband spectrum of the sub-mm data including ALMA
predominantly with the hadronic cascade
may be possible by considering a suitable radial distribution of physical properties, beyond our one zone model,
to be studied in the future.

\section{Additional caveats}

To account for the observed soft neutrino spectrum at TeV energies with our model,
the parameter $\eta_g$ (sometimes called the ``gyrofactor'' or ``Bohm factor'')
describing the strength of magnetic turbulence at the acceleration site
must be $\sim 10^3-10^4$.
Although physically possible, there is no obvious reason
for such values to be preferentially realized in the conditions discussed here.
Observationally, 
even larger values of $\eta_g \gtrsim 10^4$ are inferred for electron acceleration in blazars,
but the underlying reason is not well understood \cite{Inoue96}.
On the other hand, observations of supernova remnants reveal $\eta_g \sim 1-40$ \cite{Tsuji21},
closer to the ``Bohm limit'' of $\eta_g = 1$ corresponding to fully turbulent magnetic fields on the scales of particle gyroradii,
for which there are some physical grounds based on instabilities induced by the particles themselves \cite{Bell13}.

Our model may still be compatible with such lower values of $\eta_g$
if we consider the acceleration site to be separate and at appreciably larger $R$ than the emission region with $R = 30 R_s$.
For example, $R= 300 R_s$ can be consistent with regions of failed, line-driven winds that potentially harbor shocks \cite{Sim10}.
As $t_{\rm acc} \propto R^{5/2}$, $t_{\rm dyn} \propto R^{3/2}$ and $t_{pp} \propto R^{1/2}$,
at $R= 300 R_s$, the maximum proton energy is limited by $pp$ losses to
$E_{p,\max} \sim 200 {\rm \, TeV} (\eta_g/30)^{-1} (\epsilon_B/0.1)^{1/2} (R/R_s)^{-2}$,
with fiducial values for $L_p$ and $\epsilon_p$.
If a sizable fraction of these accelerated particles can then be transported inward advectively to $R = 30 R_s$
and induce the processes discussed in the main text,
the observed MM emission may be reproduced with relatively low values of $\eta_g$.
More discussion of such multi-zone models are deferred to future work.

Our formulation is based on the assumption that
the circulating flows in the inner regions of failed line-driven winds induces an ensemble of shocks and DSA of protons.
However,
it is possible that the actual flow in failed winds is closer to turbulence.
If so, particle acceleration may still occur via mechanisms such as
stochastic acceleration or magnetic reconnection,
as envisaged in some corona models \cite{Murase20, Kheirandish21}.
The resulting neutrino spectrum would then depend on the properties of the relevant turbulence,
which is quite uncertain.

As an alternative to line-driven mechanisms,
AGN winds may form primarily by magnetohydrodynamic processes \cite{Blandford82, Fukumura17, Kazanas19},
in which case more coherent outflows may be expected compared to failed, line-driven winds.
For dissipation of the wind kinetic energy 
and particle acceleration to occur at $R \lesssim 100 R_s$,
some extrinsic process is required, e.g. interaction with ambient BLR clouds \cite{Moriya17, Inoue22}.
In this case, it is not obvious whether shocks can result with the properties required for our model. 

Similar considerations apply to the jet known to exist in NGC 1068,
extending to $R \sim$ 0.8 kpc and aligned with the NLR outflow
\cite{Wilson87, Gallimore96_radio2},
with estimated velocity $v_j \sim 0.04 c$ \cite{Roy00}
and power $L_j \sim 2 \times 10^{43} {\rm \, erg/s}$ \cite{GarciaBurillo14},
much smaller and weaker than the jets of radio-loud AGN \cite{Blandford19}.
In principle, protons may be accelerated therein \cite{Alvarez04, Peer09},
but it is unclear if the conditions discussed above can be met.

For the outer region, our discussion focused only on $pp$ processes with simplifying assumptions.
More realistically,
1) the torus gas is likely clumpy and the structure of the wind-torus interface can be non-trivial \cite{Mou21a},
2) synchrotron and IC emission from primary electrons can dominate \cite{Mou21b}, and
3) a starburst contribution to the GeV-TeV emission can be non-negligible \cite{FermiLAT12, YoastHull14, MW16, Eichmann16}.
These aspects will be considered in future work.

\section{Additional tests}

We discuss additional tests of the model.
Variations in the mass accretion rate ${\dot M}$ 
will be reflected in those of the disk plus corona luminosity
as well as the wind power, especially if the latter is radiatively driven.
This can cause correlated variability between neutrinos, sub-GeV, sub-mm and optical-X-ray emission,
which may be observationally tractable on timescales of months to years.
Even when the latter is largely obscured as in NGC 1068, it can still be probed to some extent through the polarized IR-UV emission \cite{Marin18}.
However, these expectations may also apply to the coronal region models.
Possible evidence for year-timescale variability of neutrinos from NGC 1068 has been presented \cite{IceCube21}.

Gamma rays $\gtrsim 1$ GeV may also exhibit some variability due to changes in the clumpy structure of the wind-torus interface on timescales $t_{\rm dyn,o} \lesssim$ 20 yr.
In contrast, virtually no variability is expected if their origin is the wind external shock or the host galaxy.
Some correlation with neutrinos due to ${\dot M}$ variations may also occur, but probably only on timescales $\sim R_{\rm tor}/v_o \gtrsim$ 60 yr.

The conspicuous cascade emission in the sub-millimeter band is a unique feature of our model
and a potentially critical discriminant from the coronal region models.
Compared to $B \sim 100 {\rm \, G}$ for our inner region,
the disk corona model entails much stronger magnetic fields $\sim 1000 {\rm \, G}$ \cite{Murase20, Kheirandish21},
so that the cascade may not extend below IR frequencies due to stronger SSA.
On the other hand, the accretion shock model invokes much weaker fields $\sim 10 {\rm \, G}$ \cite{YInoue19, YInoue20},
implying much less luminous cascade; in fact, primary electrons are assumed to dominate the synchrotron emission.
At sub-mm frequencies, other emission components could be important as well,
such as the dusty torus, pc-scale jet and ambient ionized gas \cite{Michiyama23}.
The cascade emission originating from $R \sim 30 R_s$ can be 
distentangled from such components arising from larger scales
through variability and MM correlations as described above.
Further tests may be provided by VLBI imaging, e.g. by ngVLA that can reach angular resolution $\lesssim 0.1$ milli-arcsec
at $\nu \sim$ 100 GHz \cite{Reid18},
where the cascade emission from $\sim$3 micro-arcsec should remain unresolved, as opposed to other components from larger scales.
Realistic modeling of the sub-mm emission from NGC 1068 including all emission components will be presented in the future.

As a candidate source other than NGC 1068, 
deeper studies are particularly warranted for the Circinus galaxy, the nearest Seyfert galaxy at $D \sim$ 4 Mpc,
with $M_{\rm BH} \simeq 2 \times 10^6 M_\odot$ and $\lambda_{\rm Edd} \simeq 0.2$ \cite{Greenhill03},
harboring a prominent wind \cite{Wilson00}.
Although neutrino detection may need to await KM3NeT \cite{KM3NeT16}
due to the source's southern location \cite{Kheirandish21},
its GeV gamma-rays are of unknown nature,
with luminosity possibly in excess of the star-formation contribution \cite{Hayashida13},
and marginal evidence of variability \cite{Gou19}.

A search for neutrinos from radio-quiet AGN was conducted by IceCube
using a catalog of soft X-ray selected objects combined with mid-infrared or radio catalogs \cite{IceCube22AGNcore}.
Studies based on X-rays are inevitably biased toward unobscured type-1 AGN and against obscured type-2 AGN,
even though the neutrino emission is not expected to depend much on AGN orientation.
A more recent work utilized hard X-ray selected Seyfert AGN
where the intrinsic X-ray flux before obscuration had been estimated for each object \cite{IceCube24Seyfert}.
Although less severe than for soft X-rays,
attenuation effects can still be significant for hard X-rays,
and estimating the intrinsic flux entail large uncertainties,
especially for Compton-thick objects that constitute a sizable fraction of all AGN \cite{RamosAlmeida17}.
For the well-studied case of NGC 1068, Ref. \cite{Bauer15} found that the data including hard X-rays
cannot be explained by a simple model of obscuration but required
three different types of absorbers with different column densities and spatial distributions,
some of which is also known to be time variable \cite{Marinucci16, Zaino20}.
In this regard, an alternative strategy may to employ a sample of radio-quiet AGN
selected by their mid-infrared emission alone.
Unlike X-rays, such emission provides a reasonable measure of the total radiative output due to BH accretion
that is relatively free of orientation effects.

Besides non-thermal MM emission,
the high density of energetic protons around the nucleus implied here
may lead to other observable signatures,
e.g. characteristic effects on the molecular chemistry of the torus \cite{Aladro13},
which is worth investigating.
For the wind-torus interaction,
thermal and/or kinematic signatures in emission lines characteristic to shocks \cite{DAgostino19}
would be valuable,
although separating the contribution from the inner torus region may be challenging.

\section{Comparison with other studies}

Early models of neutrino emission from AGN cores
were based on speculative mechanisms for proton acceleration when the basic nature of AGN was less known compared to today \cite{Eichler79, Berezinsky81}.
Later models
postulated proton acceleration in accretion shocks with very high efficiency \cite{Kazanas86, Begelman90, Stecker91},
at a time when non-thermal pair cascades were considered promising as the origin of the X-ray emission from radio-quiet AGN. 
However, such models have been disfavored since the detection of hard X-ray spectral cutoffs instead of prominent electron-positron annihilation features in nearby Seyfert galaxies,
which strongly supported thermal Comptonization as the X-ray emission mechanism \cite{YInoue21,MS22}.

In view of the predominantly thermal nature of X-ray emission in radio-quiet AGN,
more recent models of neutrino emission invoke proton acceleration with more moderate efficiency
in hot coronal regions near the BH, 
either accretion disk coronae \cite{Murase20, Kheirandish21} or accretion shocks \cite{YInoue19, YInoue20, Anchordoqui21}.
However, despite indications from test particle simulations of stochastic acceleration \cite{Kimura16, Kimura19b}
and radio to sub-mm observations of some nearby AGN \cite{Panessa19, YInoue21},
acceleration of non-thermal particles in such coronal regions is not yet unequivocally established from either observations or theory.
Some proposed mechanisms such as stochastic acceleration and magnetic reconnection entail large uncertainties \cite{Kheirandish21}.
In such regions, $\gamma\gamma$ absorption is effective down to the MeV range, so the observed gamma rays at $\gtrsim$ GeV must arise from a separate region.
These studies also did not discuss the potential contribution of electromagnetic (EM) emission from the consequent electron-positron pair cascade down to the radio band.
Some of these issues were addressed here in the context of DSA in the inner regions of AGN disk-driven winds.

The small-scale jets of radio-quiet AGN have also been discussed as sites of high-energy emission,
e.g. $p\gamma$ neutrino emission by accelerated protons \cite{Alvarez04},
and IC emission by accelerated electrons to account for the GeV gamma rays from NGC 1068 \cite{Lenain10}.

For AGN winds, theoretical studies have addressed the observability of non-thermal emission from external shocks
where the winds interact with the host galaxy gas
\cite{Faucher12, Tamborra14, Nims15, Wang15, Lamastra16, Liu18}.
Tentative evidence of such emission in the radio band has been presented \cite{Zakamska14, Panessa19, Richards21},
but are not yet conclusive.
Searches for positional correlations between IceCube neutrinos and AGN with prominent winds have so far been negative \cite{Padovani18}.

Evidence was recently presented for GeV gamma-ray emission in a stacked sample of AGN in which UFOs have been detected, and proposed as $pp$ gamma-rays from external shocks induced by AGN winds in the host galaxy interstellar medium (ISM) \cite{Fermi-LAT21}.
The reality of the association between the gamma-ray emission and AGN winds can be questioned, since currently available samples of UFOs are subject to unknown biases concerning their viewing angle dependence \cite{Nomura16, Giustini19} and time variability \cite{Igo20}. 
Even if the connection between GeV emission and AGN winds is real,
the above interpretation may be problematic, as the propagation of winds from the sub-pc scales of UFOs to the kpc scales of the host galaxy entails significant time delays, $\gtrsim 10^5$ yr.
Ref. \cite{Peretti23} proposed that the Fermi results can be explained by AGN wind external shocks occurring on pc scales, but their assumption of an ambient medium with constant density $\sim 10^4 {\rm \, cm^{-3}} $ is quite ad-hoc and does not adequately reflect the actual sub-pc environment of AGN that includes the ubiquitous dusty torus.
In this respect,
the GeV emission mechanisms discussed here induced by failed winds and wind-torus interaction may be more viable.

Non-thermal EM emission assuming DSA in shocks due to collisions between BLR clouds and the accretion disk
has been proposed \cite{Mueller20}.
However, for relatively high $\lambda_{\rm Edd}$ objects like NGC 1068,
it is unclear whether BLR clouds can impact the disk at all;
more likely they will be pushed out by the disk radiation pressure.
Moreover, even if the impact can somehow occur with sufficient velocity,
the disk gas density can be high enough so that the resulting shock is collisional, suppressing DSA.

As this work was being completed, a number of studies have appeared in the literature as reviewed in Ref. \cite{Padovani24}.

\end{document}